\documentclass[english,a4paper]{article}

\usepackage{graphicx}
\usepackage{amsmath}
\usepackage{amssymb}
\usepackage{mathtools}
\usepackage{color}

\def\phi{\varphi}
\def\epsilon{\varepsilon}

\everymath{\displaystyle}

\newtheorem{theorem}{Theorem}

\newtheorem{lemma}[theorem]{Lemma}

\newcommand*\ve{\pmb e}

\begin{document}
\title{A greedy reconstruction algorithm for the identification of spin distribution}
\author{S. Buchwald\footnote{Universit\"at Konstanz, Universit\"atsstra\ss e 10, 78464 Konstanz, Germany}, G. Ciaramella\footnote{MOX-Laboratory, Dipartimento di Matematica, Politecnico di Milano, Piazza Leonardo da Vinci 32, 20133 Milano, Italy}, J. Salomon\footnote{INRIA Paris, ANGE team, Building A, office 323, 2 rue Simone Iff, 75589 Paris, France}, D. Sugny\footnote{Laboratoire Interdisciplinaire Carnot de Bourgogne (ICB), UMR 6303 CNRS-Universit\'e Bourgogne-Franche Comt\'e, 9 Av. A.
Savary, BP 47 870, F-21078 Dijon Cedex, France; dominique.sugny@u-bourgogne.fr}}

\maketitle

\begin{abstract}
We propose a greedy reconstruction algorithm to find the probability distribution of a parameter characterizing an inhomogeneous spin ensemble. The identification is based on the application of a
number of constant control processes during a given time for which the final ensemble magnetization vector is measured. From these experimental data, we show that the
identifiability of a piecewise constant approximation of the probability distribution is related to the invertibility of a matrix
which depends on the different control protocols applied to the system. The algorithm aims to design specific controls which ensure that this matrix is as
far as possible from a singular matrix. Numerical simulations reveal the efficiency of this algorithm in different examples. A systematic comparison with respect to random  constant pulses is done.
\end{abstract}


\section{Introduction}\label{sec1}
The identification of parameters that characterize the dynamics of a quantum system is a fundamental prerequisite for controlling
its evolution~\cite{brif:2010,glaser_training_2015,RMP:rotation,koch_controlling_2016,BCSbook,monschemes,sprengel2018investigation,BSS} and is
of practical interest for realizing specific tasks in quantum technologies~\cite{QT}. This aspect is crucial in open-loop configurations for which
the control protocols are designed without any experimental feedback from the system during the control process~\cite{brif:2010,glaser_training_2015,alessandrobook,brysonbook,BCSbook}.
In the context of quantum systems, the problem of identifying unknown parameters (or functions) has been explored by a large number of studies and for a variety of applications ranging from molecular
physics~\cite{geremia2003,feng2006,geremia2002} and magnetic resonance~\cite{ma2013,ansel2017,pierre2016,CBDW2015} to quantum information
science~\cite{schirmer2004,schirmer2009,schirmer2015,yuan2017,liu2017,wittler2021,zhang2014,sone2017,kiukas2017,burgarth2017,Wang_2015} and open quantum systems~\cite{xue2021,xue2020,zhang2015,Bennink_2019}. Some mathematical results have also been established
in this direction~\cite{ljung2010,lebris06,alis2004,ndong2014,rojas2007,baudouin2008,bonnabel2009,fu2017,wang2018}. On the basis of different measurement processes and specific control protocols, the goal of these works is generally to estimate the value of one or
several parameters of the system Hamiltonian. When controlling an ensemble of identical quantum systems, such a parameter may vary
in a given range due to experimental limitations or uncertainties. A key example comes from the spatial inhomogeneities of the external
control~\cite{Li06,li09,glaser_training_2015,kobzar:2008,lapert_exploring_2012,turinici2019}. In this case, all the systems are not subjected exactly to the same control. This aspect has to be taken into account in the modeling of the dynamics and in the computation of the control procedure. Robust control
protocols against such inaccuracies have been developed recently to solve this experimental issue~\cite{glaser_training_2015,kobzar:2008,vandamme2017,Ruschhaupt_2012,vandamme2017_2,daems2013,genov2014,barnes1,barnes2,barnes3}.
However, the variation range of the unknown parameter is not the only crucial quantity, the probability distribution of this parameter (i.e. the number
of systems for each value of the parameter) may play also a major role. It is generally assumed that this probability distribution is flat or has a simple Gaussian
or Lorentzian form. In these cases, the probability distribution can be quite easily characterized. 
However, the problem of identifying probability distributions becomes much more difficult if these have complex structures, with e.g. several peaks, or if no information is known about them. It is therefore essential to be able to identify with a great precision
these unknown probability distributions.

This paper aims at taking a step toward the answer to this open question by developing a numerical algorithm, called a Greedy Reconstruction Algorithm (GRA). By definition, an algorithm
is said to be greedy if it takes the best choice available at each iterative step. Greedy algorithms find generally a sub-optimal solution, but in a computational time which may be very
small compared to the one of a global optimization procedure. Such algorithms have been recently applied to the identification of quantum systems~\cite{madaysalomon,BCS2021} and we propose
to adapt them to the reconstruction of probability distribution. For the sake of clarity, we focus in this study on a specific example, although our algorithm applies to a large variety of systems. We consider the case of a spin ensemble in Nuclear Magnetic Resonance (NMR)~\cite{li09,conolly:1986,lapertprl,Levitt08,skinner:2005} subjected
to an inhomogeneous radio-frequency magnetic field whose range of variation is known, but not its probability distribution. The probability distribution is approximated by a piecewise constant function taking at most $K$ values. The algorithm then designs a series of $K$ controls for GRA (or less for the optimized version) which are, in a second step, applied to the spin ensemble.
The $K$ measured ensemble magnetization vectors at the final time are then used to identify the probability distribution. More precisely, the identification process is related to
the invertibility of a matrix which depends on the different controls. The aim of the algorithm is therefore to design specific control protocols which ensure that this matrix is as
far as possible from being singular. The precision of the identification process can be understood from the eigenvalues and eigenvectors of the matrix. In the examples analyzed in this study, constant controls will be sufficient to find the probability distribution with a very good accuracy, but time-dependent controls could also be used. We show that the optimization procedure of the algorithm has a unique solution and a good convexity structure leading to fast convergence.  We point out that the controls only depend on the model system and not on the spin distribution or on the available data. The numerical efficiency of the algorithms is shown on different illustrative examples, namely a double-peak distribution and a step one. A systematic comparison with random constant pulses is also done.

The paper is organized as follows. Section~\ref{sec2} describes the model system. Section~\ref{sec:distribution} is dedicated to the theoretical framework,
while the algorithm is presented in Sec.~\ref{secalgo}. The efficiency of GRA is numerically demonstrated in Sec.~\ref{secnum} on two standard examples, a double-peak and a step probability distributions. Conclusion and prospective views are given in Sec.~\ref{sec:conclusions}. A mathematical description of the method and the development of an optimized greedy algorithm are reported respectively in Appendices~\ref{mathdesc} and~\ref{OGRA}. Additional numerical results are presented in the supplemental material~\cite{suppmat}.
\section{The model system}\label{sec2}
To illustrate our study, we consider a basic control problem in NMR, i.e. a spin ensemble subjected to an inhomogeneous radio-frequency magnetic field~\cite{Li06,li09,kobzar:2008,skinner:2005,lapert_exploring_2012}. In a given rotating frame, we assume that all the spins have the same resonance offset $\omega$. Each isochromat is characterized by a Bloch vector $\textbf{M}(\alpha)=[M_x,M_y,M_z]^\intercal$ whose dynamics are governed by the following equations of motion:
$$
\begin{cases}
\dot{M}_x=-\omega M_y+(1+\alpha)\omega_yM_z, \\
\dot{M}_y=\omega M_x-(1+\alpha)\omega_xM_z, \\
\dot{M}_z=(1+\alpha)\omega_xM_y-(1+\alpha)\omega_yM_z,
\end{cases}
$$
where the coordinates of the Bloch vector satisfy $M_x^2+M_y^2+M_z^2=M_0^2$, with $M_0$ the equilibrium magnetization. $\omega_x$ and $\omega_y$ are time-dependent controls that  correspond to the components of the magnetic field along the $x$- and the $y$- directions. In this study, we assume that these controls are constant in time. We show in Sec.~\ref{secnum} that this hypothesis is sufficient for the different examples to identify the probability distributions. The parameter $\alpha$ is used to model the control field inhomogeneities which are of the order of few percents in standard experiments~\cite{Levitt08}. The controls $\frac{\omega_x}{2\pi}$ and $\frac{\omega_y}{2\pi}$ are expressed in Hz. We consider a typical field amplitude $\omega_0$ that can be fixed, for instance, to $\omega_0=2\pi\times 100$~Hz. We introduce normalized coordinates as follows:
$$
u_x=2\pi\frac{\omega_x}{\omega_0};~u_y=2\pi\frac{\omega_y}{\omega_0};~t'=\frac{\omega_0}{2\pi}t;~\Delta=2\pi\frac{\omega}{\omega_0};\textbf{X}=\frac{\textbf{M}}{M_0}.
$$
We omit the $'$ in the time below to simplify the notations. We deduce that the differential system can be expressed in normalized units as:
\begin{equation}\label{eq1}
\begin{cases}
\dot{x}=-\Delta y+(1+\alpha)u_yz \\
\dot{y}=\Delta x-(1+\alpha)u_xz \\
\dot{z}=(1+\alpha)u_xy-(1+\alpha)u_yz
\end{cases}
\end{equation}
with $x^2+y^2+z^2=1$. The initial state of the dynamics for each spin is the thermal equilibrium point, i.e. the north pole of the Bloch sphere, $\textbf{X}_0=(0,0,1)^\intercal $. We neglect the relaxation effect and we consider a control time of the order of 100~ms. This corresponds to a normalized time $t_f$ of the order of 10. In the numerical simulations, we add the constraints $|u_x|\leq u_m$ and $|u_y|\leq u_m$ where $u_m$ is the maximum amplitude of each component.  In NMR, only the first two coordinates of the magnetization vector can be directly measured. We do not have accessed to the $z$- component due to the strong magnetic field applied along this direction. We denote by $\textbf{Y}(t)=(x(t),y(t))^\intercal$ the projection of the Bloch vector onto the first two coordinates. We point out that this aspect is not a limiting point for the application of the identification process.
\section{Identification of spin distribution}\label{sec:distribution}
We consider an ensemble of $N$ spins whose dynamics are governed by Eq.~\eqref{eq1}. We assume that the control amplitudes $(u_x,u_y)$ belong to the admissible set $\mathcal{U}=\{(u_x,u_y)\in\mathbb{R}^2\mid|u_x|\leq u_m, |u_y|\leq u_m \}$. The objective of the control procedure is to identify the probability distribution of the parameter $\alpha$. To simplify the recognition process, we assume that the ensemble of spins can be decomposed into a set of $K$ subgroups with the same value of the parameter $\alpha_\ell$, $1\leq \ell \leq K$. However, the discrete probability distribution $P_\star$ for $\alpha$, namely the probability of each possible outcome $\alpha_\ell$, or in other words the number of elements $N_\ell$ of each subgroup, is unknown. By definition, we have $P_\star(\ell)=\frac{N_\ell}{N}$ and $\sum_{\ell=1}^{K}P_\star(\ell)=1$.

The projected solution onto the first two coordinates at time $t_f$ of Eq.~\eqref{eq1} is denoted by $\textbf{Y}_{\textbf{u},\alpha}(t_f)$ where the dependance on $\textbf{u}$ and $\alpha$ has been explicitly mentioned. The corresponding experimental realization of this controlled dynamic leads to $\textbf{Y}^{\textrm{exp}}_{\textbf{u}}(t_f)=(x^{\textrm{exp}}_{\textbf{u}}(t_f),y^{\textrm{exp}}_{\textbf{u}}(t_f))^\intercal$, where $\textbf{Y}^{\textrm{exp}}_{\textbf{u}}(t_f)$ can be viewed as the average at time $t_f$ of the experimental measures of all the spins of the set subjected to the control $\textbf{u}$. The coordinates $x^{\textrm{exp}}_{\textbf{u}}$ and $y^{\textrm{exp}}_{\textbf{u}}$ are the ones of this measured magnetization vector.

The relation between the theoretical description of the dynamical system to the experimental outcome can be expressed as:
\begin{equation}\label{eqpstar}
\textbf{Y}^{\textrm{exp}}_{\textbf{u}}(t_f)=\sum_{\ell=1}^{K}P_\star(\ell)\textbf{Y}_{\textbf{u},\alpha_\ell}(t_f),
\end{equation}
in which the two sides of the equation crucially depend on the control $\textbf{u}$. A specific control protocol is not sufficient to identify the probability distribution $P_\star$ which generally requires the implementation of $K$ control processes with $K$ different controls denoted $\textbf{u}_k$, $k=1,\cdots,K$. Note that in the optimized version of the GRA presented in Appendix~\ref{OGRA}, the number of controls can be different from $K$.

On the basis of the experimental outputs, a straightforward way to determine $P_\star$ is to solve the following minimization problem:
\begin{equation}\label{pbmin}
\min_{P\in \mathbb{P}}\sum_{k=1}^{K}\|\textbf{Y}^{\textrm{exp}}_{\textbf{u}_k}(t_f)-\sum_{\ell=1}^{K}P(\ell)\textbf{Y}_{\textbf{u}_k,\alpha_\ell}(t_f)\|^2,
\end{equation}
where $\mathbb{P}$ is the set of all the possible probability distributions $P$ that satisfy $P(\ell)\geq 0$ for $1\leq \ell\leq K$ and $\sum_{\ell=1}^{K}P(\ell)=1$. Mathematically, we point out that $\mathbb{P}$ is a convex and closed set. $\|\cdot\|$ denotes the standard Euclidean vector norm. Note that Eq.~\eqref{pbmin} can be rewritten as:
\begin{equation}\label{pbmin2}
\min_{P\in \mathbb{P}}\sum_{k=1}^{K}\|\sum_{\ell=1}^{K}(P_\star(\ell)-P(\ell))\textbf{Y}_{\textbf{u}_k,\alpha_\ell}(t_f)\|^2 .
\end{equation}
At this point, it is clear that a key ingredient of the accuracy of the identification process relies on the choice of the controls $\textbf{u}_k$.

To clarify this problem, we introduce a set $\{\phi_j\}_{j=1}^{K}$ of linearly independent functions $\phi_j:\{1,\cdots,K\}\to \mathbb{R}$ such that $\mathbb{P} \subset {\rm span}(\{\phi_j\}_{j=1}^{K})$, where $\textrm{span}$ denotes the vector space generated by the functions. Expressing respectively $P_\star$ and $P$ as $P_\star(\ell)=\sum_{j=1}^{K}\beta_{\star,j}\phi_j(\ell)$ and $P(\ell)=\sum_{j=1}^{K}\beta_{j}\phi_j(\ell)$, the minimization problem \eqref{pbmin2} becomes:
\begin{equation}\label{eqbeta}
\min_{\beta\in \widehat{\mathbb{R}}^K}\sum_{k=1}^{K}\|\sum_{\ell,j=1}^{K}(\beta_{\star,j}-\beta_j)\phi_j(\ell)\textbf{Y}_{\textbf{u}_k,\alpha_\ell}(t_f)\|^2,
\end{equation}
where the vector $\beta=(\beta_j)_{j=1}^K$ is taken in $\widehat{\mathbb{R}}^K$, a subset of $\mathbb{R}^K$, so that $P=\sum_j\beta_j\phi_j$ is a probability distribution. Equation~\eqref{eqbeta} can be rewritten in a compact form as follows:
\begin{equation}\label{eq: pbminW}
\min_{\beta\in \widehat{\mathbb{R}}^K}\langle \beta_\star-\beta|W|\beta_\star-\beta\rangle,
\end{equation}
where $W$ is a symmetric and positive semi-definite $K\times K$- matrix whose elements are defined as:
\begin{equation}\label{eq: W}
W_{\ell,j}=\sum_k\langle \gamma_{\ell}(\textbf{u}_k)|\gamma_j(\textbf{u}_k)\rangle
\end{equation}
with
$$
\gamma_j(\textbf{u}_k)=\sum_\ell\phi_j(\ell)\textbf{Y}_{\textbf{u}_k,\alpha_\ell}(t_f).
$$
Since the set of vectors $\beta$ is a convex subset of $\mathbb{R}^{K}$, we deduce that the problem is uniquely solvable if the matrix $W$ is positive definite, i.e. if $W$ has a non-zero determinant. In the case $W$ has a non-trivial kernel, infinitely many solutions may exist which lead to wrong probability distributions different from the experimental one $P_\star$. We stress that the non-triviality of the kernel depends completely on the choice of the controls $\textbf{u}_k$.

We show in this study that GRA allows us to design a set of controls $\textbf{u}_k$ so that the matrix $W$ is positive definite with a trivial kernel. The algorithm is composed of two steps, namely an offline and an online steps. In the first stage, GRA computes the controls $\textbf{u}_k$.
In this phase, only the theoretical model is needed without any experimental input. The derived controls are used in a second step in which the different
magnetization vectors are measured and the minimization problem~\eqref{pbmin} is solved. Note that the controls are the same for any probability distribution
to identify and only depend on the model system under study. Finally, we point out that, while in a first algorithm we consider that all control pulses have the same duration
$t_f$, in a second version described in Sec.~\ref{sectimeopt}, the duration of each pulse is considered as a variable to be optimized together with its amplitude. The generality of GRA allows one to tackle this situation in a straightforward manner.

\section{A greedy reconstruction algorithm}\label{secalgo}
We present in this section the GRA in its classical form, an optimized extension called optimized GRA (OGRA) is described in Appendix~\ref{OGRA}. For pedagogical purposes, we have limited the mathematical derivation of the algorithm to its strict minimum. The interested reader can find mathematical details about the algorithms in~\cite{madaysalomon} and \cite{BCS2021} for the standard and optimized GRA, respectively.
\subsection{Optimizing the control amplitudes for a fixed control time}
The GRA computes the controls $\textbf{u}_k$ by solving a sequence of fitting-step and discriminatory-step problems, in which the goal of the first step is to identify a nontrivial kernel of a sub-matrix of $W$, while the second phase designs a new control which is aimed to correct this discrepancy and to eliminate the identified non-trivial kernel. The explicit formulation of the algorithm is given in terms of the function $\textbf{h}^{(k)}$ defined by:
\begin{equation}\label{hk}
\textbf{h}^{(k)}(\beta,\textbf{u})=\sum_{\ell=1}^{K}\sum_{j=1}^k\beta_j\phi_j(\ell)\textbf{Y}_{\textbf{u},\alpha_\ell}(t_f),
\end{equation}
for any $\beta$ in $\mathbb{R}^k$.
GRA is described below. Some mathematical statements of the different steps of the algorithm are described in Appendix~\ref{mathdesc}. Its numerical implementation is presented and discussed in Sec.~\ref{secnum}.\\ \\
\noindent {\bf Greedy Reconstruction Algorithm (GRA):} Given a set of $K$ linearly independent functions $(\phi_1,\ldots,\phi_{K})$.\\
Solve the initialization problem
\begin{equation}\label{eq: initialization}
\max_{\textbf{u}\in\mathcal{U}} \|\textbf{h}^{(1)}(1,\textbf{u})\|^2,
\end{equation}
that gives the control $\textbf{u}_1$, and set $k=1$.\\
\textbf{While} $k\leq K-1$
\begin{enumerate}
\item \underline{Fitting step}: Find $(\beta^{k}_j)_{j=1,\dots,k}$ that solves the problem
		\begin{equation}\label{eq: fitting step}
		\min_{\beta \in \mathbb{R}^k}\sum_{m=1}^{k}\| \textbf{h}^{(K)}(\textbf{e}_{k+1},\textbf{u}_m)-\textbf{h}^{(k)}(\beta,\textbf{u}_m)\|^2,
		\end{equation}
where $\textbf{e}_{k+1}$ is the $(k+1)$-th canonical vector in $\mathbb{R}^K$.
\item \underline{Discriminatory step}: Find $\textbf{u}_{k+1}$ that solves the problem
		\begin{equation}\label{eq: discriminatory step}
		\max_{\textbf{u}\in\mathcal{U}}\|\textbf{h}^{(K)}(\textbf{e}_{k+1},\textbf{u})-\textbf{h}^{(k)}(\beta_k,\textbf{u})\|^2.
		\end{equation}
\item Update $k+1 \rightarrow k$.
\end{enumerate}
\textbf{End while}\\ \\
The basic principles of GRA can be detailed by its two first iterations for $K=2$. Using
$$
\textbf{h}^{(1)}(1,\textbf{u})=\sum_{\ell=1}^K\phi_1(\ell)\textbf{Y}_{\textbf{u},\alpha_\ell}(t_f),
$$
the initialization problem can be expressed as:
$$
\max_{\textbf{u}\in\mathcal{U}} \|W_{11}(\textbf{u})\|^2.
$$
We deduce that the goal of this step is to maximize the modulus of this $W$- matrix element, so as to be as far as possible from a zero of $W$. We then consider the first step of the algorithm with $k=1$. We omit below for clarity the dependence on $\textbf{u}$ of $W$. By definition, we have:
$$
\begin{cases}
W_{11}=\|\sum_{\ell=1}^K\phi_1(\ell)\textbf{Y}_{\textbf{u},\alpha_\ell}(t_f)\|^2 \\
W_{22}=\|\sum_{\ell=1}^K\phi_2(\ell)\textbf{Y}_{\textbf{u},\alpha_\ell}(t_f)\|^2 \\
W_{12}=W_{21}=\sum_{\ell,\ell'=1}^K\phi_1(\ell)\phi_2(\ell')\langle \textbf{Y}_{\textbf{u},\alpha_\ell}(t_f)|\textbf{Y}_{\textbf{u},\alpha_{\ell'}}(t_f)\rangle
\end{cases}
$$
and we deduce that the quantity to minimize in the fitting step can be written as:
$$
\|\textbf{h}^{(2)}(e_2,\textbf{u}_1)-\textbf{h}^{(1)}(\beta,\textbf{u}_1)\|^2=W_{11}\beta^2-2W_{12}\beta+W_{22},
$$
where $\beta$ is here a real number. The minimum is reached for $\beta_1=W_{11}^{-1}W_{12}$ where $W$ is computed for the control $\textbf{u}_1$. This value can be associated to a vector $(\beta_1,-1)^\intercal$ of the kernel of the following $2\times 2$- submatrix of W:
\begin{equation}\label{eqsubmat}
\begin{pmatrix}
W_{11} & W_{12} \\
W_{12} & W_{22}
\end{pmatrix}
\end{equation}
The fitting step of GRA can thus be interpreted as a systematic way to find a basis of the kernel of larger and larger sub-matrices of $W$. Setting $\beta$ to
$\beta_1$, the discriminatory step consists in adding a new control $\textbf{u}_2$ to correct this singularity, i.e.
in selecting this control such that the corresponding quantity is as far as possible from a zero. Mathematically, it can
be shown that this procedure has always a solution and that the new matrix $W^{(2)}$ has a non-trivial $2\times 2$ sub-matrix~\eqref{eqsubmat} (see Appendix~\ref{mathdesc} for details).

\subsection{Optimizing amplitude controls and time horizon}\label{sectimeopt}
Until now, we have considered a fixed control time $t_f$. However, it is also possible to consider controls with different control times, up to a fixed boundary $t_f^{\max}>0$. In this case, we also maximize with respect to time, meaning that the initialization and discriminatory step problems at iteration $k$ would change to
\begin{equation}\label{eq: occft initialization}
\max\limits_{\substack{\textbf{u}\in \mathcal{U},\\t_f\in [0,t_f^{\max}]}} \|\textbf{h}^{(1)}(1,\textbf{u};t_f)\|^2
\end{equation}
and
\begin{equation}\label{eq: occft discriminatory step}
\max\limits_{\substack{\textbf{u}\in \mathcal{U},\\t_f\in [0,t_f^{\max}]}}
\|\textbf{h}^{(K)}(\ve^{k+1},\textbf{u};t_f) - \textbf{h}^{(k)}(\beta^k,\textbf{u};t_f)\|^2,
\end{equation}
respectively. In Eq.~\eqref{eq: occft initialization} and \eqref{eq: occft discriminatory step}, the function $\textbf{h}^{(k)}$ is still defined as in \eqref{hk}, only with
the control time as an additional variable. Similarly, one can adapt the corresponding problems in OGRA, the optimized version described in Appendix~\ref{OGRA}. We denote by GRAt and OGRAt the two resulting algorithms.

\section{Numerical results}\label{secnum}
\subsection{The case of a double peak distribution}\label{secnumgra}
\begin{figure}[h]
	\centering
	\includegraphics[width = 0.65\textwidth]{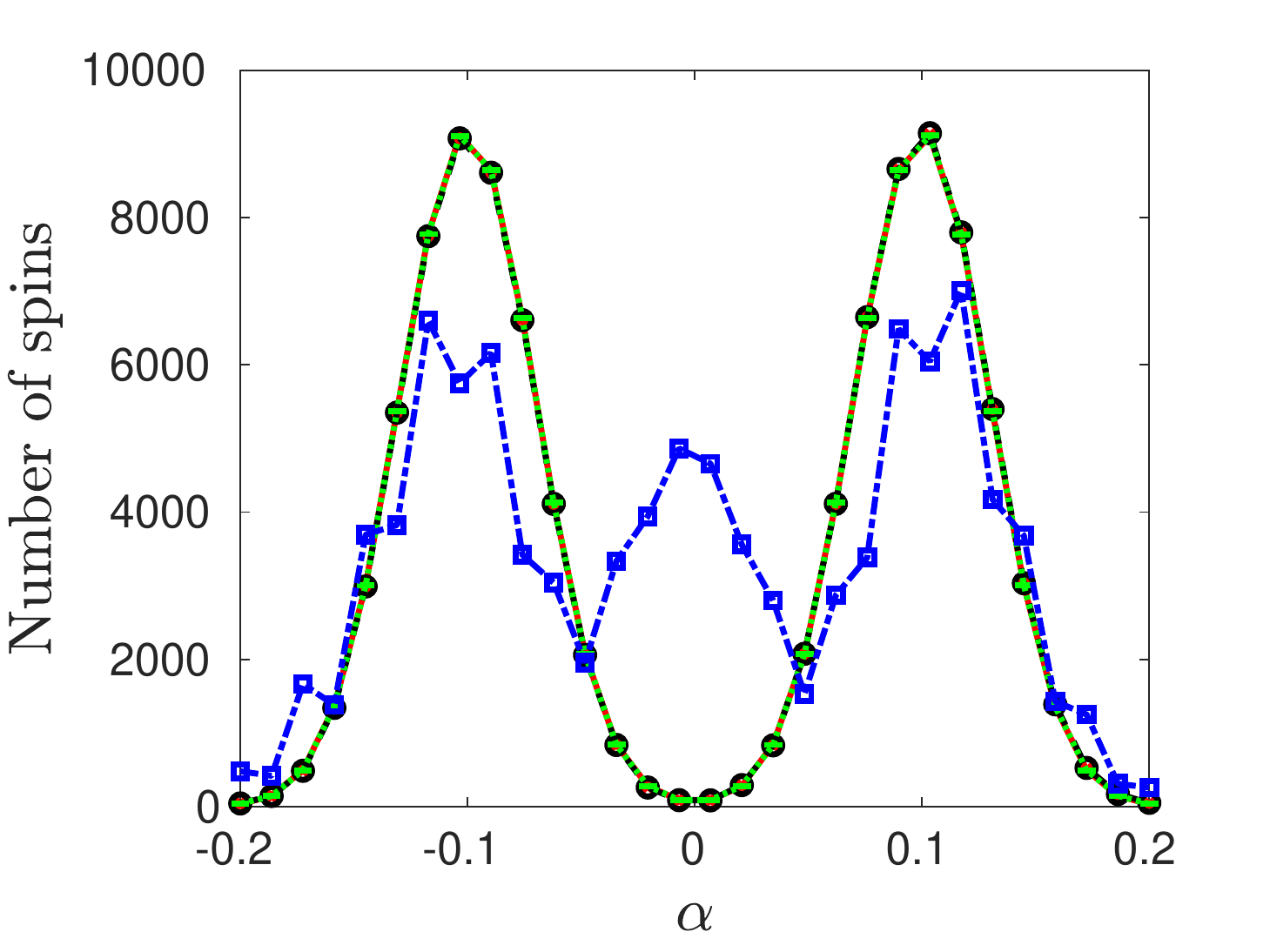}
	\caption{Plot as a function of $\alpha$ of the true distribution (red crosses) and approximated solutions with minimum error, computed by the different optimization algorithms for the identification problem using control sets with a fixed control time. In particular, controls generated by OGRA (green, vertical lines) and GRA (black, circles) and RCC (blue, squares) are plotted. Dimensionless units are used.}
	\label{fig: double peak}
\end{figure}

As a first illustrative example, we investigate in this paragraph the identification of a symmetric double peak probability distribution, displayed in Fig.~\ref{fig: double peak}. Similar results have been achieved for other smooth distributions with one or several peaks. Numerical details are described in the supplemental material~\cite{suppmat}.

In the numerical simulations, we consider a control time $t_f=t_f^{max}=16$. The amplitude $u_m$ is equal to 10. The normalized offset resonance is set for all the spins to $\frac{\pi}{10}$, i.e. to 30~Hz. We also assume that $\alpha\in [-0.2,0.2]$ and $K=30$. The $K$ discrete values $\alpha_\ell$ of $\alpha$ are regularly spaced  in the interval of variation of $\alpha$, i.e. $\alpha_\ell=-0.2+0.4\frac{\ell-1}{K-1}$. Since the control protocols are constant in time, Eq.~\eqref{eq1} is solved numerically
by directly evaluating the exponential matrix corresponding to the exact solution.
All optimization problems are solved
by a BFGS descent-direction method. We also mention that the exact number of uncoupled spins in the ensemble is not relevant for all the computations, since we are only interested in their probability distribution. However, we use a total number of $10^5$ spins in the numerical simulations.

For GRA and GRAt, we consider a random and orthonormal basis $\{\phi_k\}_{k=1}^{30}$. Note that any basis of this space can be used in the respective algorithms.
For OGRA and OGRAt, we extend the basis from GRA by 30 randomly chosen probability distributions $\{\phi_k\}_{k=31}^{60}$. The tolerance used in the OGRA and OGRAt (see Appendix \ref{OGRA}) is set to be $\textnormal{tol}=10^{-14}$. The controls generated by the algorithms and corresponding to the numerical results discussed below are described in the supplemental material~\cite{suppmat}. To test whether it is even necessary to run the algorithm or if the same results could be achieved with other control protocols, we also
consider two sets of 30 random and constant controls. For the first and second cases, we use respectively completely random constant values in the set $\mathcal{U}$, with a control time $t_f$ or with different and random control times in the interval $[0,t_f]$. We denote by RCC and RCCt the two sets of controls.

\begin{figure}[h]
	\centering
	\includegraphics[width = 0.65\textwidth]{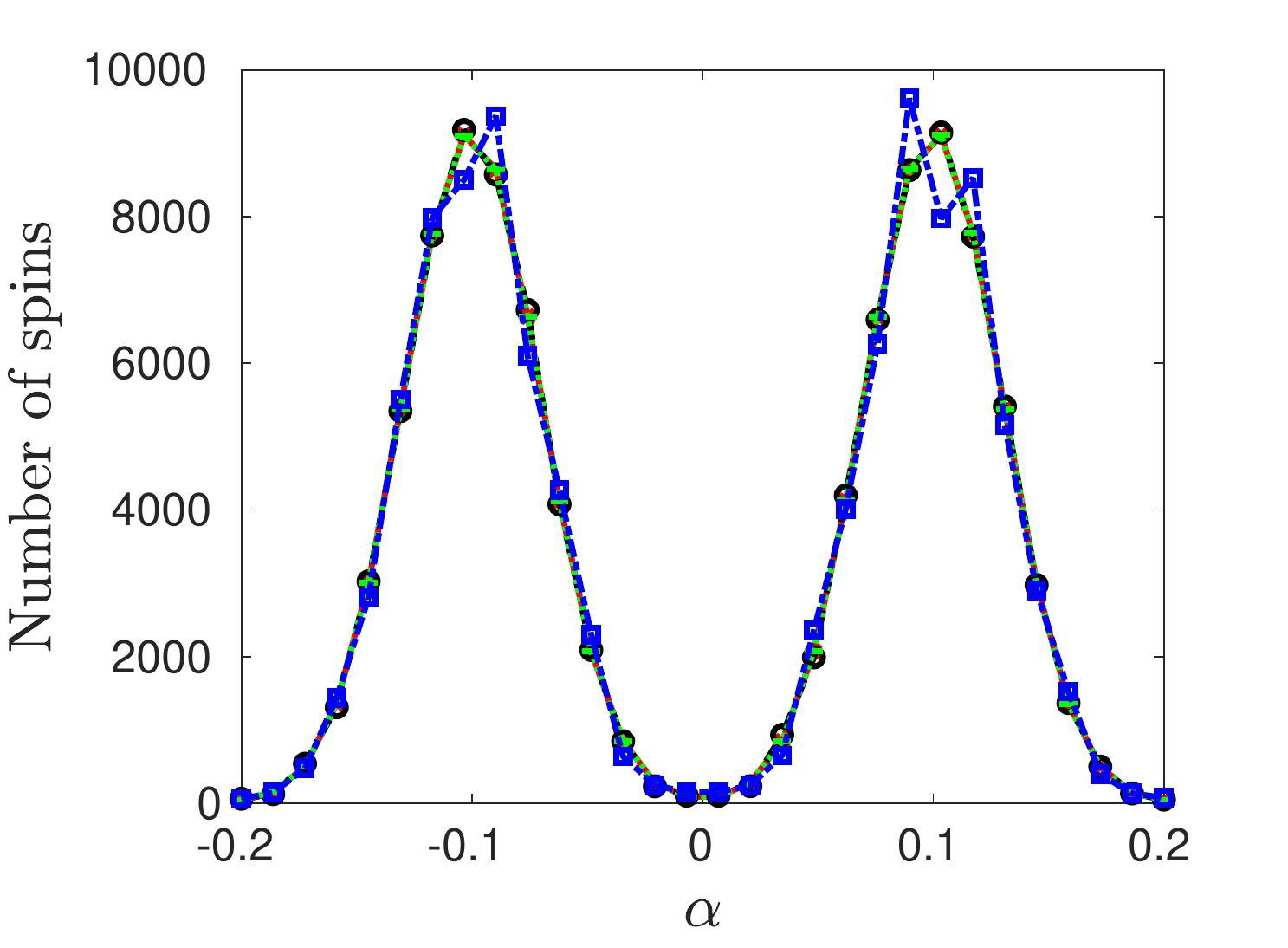}
	\caption{Same as Fig.~\ref{fig: double peak}, but for variable control times.}
	\label{fig: double peak time}
\end{figure}

The robustness of the different control functions is evaluated by considering a 30-dimensional hypercube centered in the global minimum $P_\star$ of
our identification problem, with a radius of $100\|P_\star\|$, and we repeat the minimization process for 100 initialization vectors randomly chosen in this hypercube.
We then compute the minimum norm difference $\frac{\|P_\star-P_f\|}{\|P_\star\|}$ over all optimization runs, where $P_f$ denotes the solution given by the optimization algorithm.
We obtain the results reported in Tab.~\ref{tab:1}.
\begin{table}[h]
	\centering
	\begin{tabular}{ l|c|c|c|c|c|c }
		\hline
		Control set & GRA & GRAt & OGRA & OGRAt & RCC & RCCt \\ \hline\hline
		Min. error&0.0045&0.0098&0.0005&0.0009&0.4685&0.0841\\
		\hline
	\end{tabular}
	\caption{Minimum relative norm error for different control sets.}
	\label{tab:1}
\end{table}
As can be seen in Tab.~\ref{tab:1}, the errors of OGRA and OGRAt are ten times smaller than the ones of GRA and GRAt, which themselves are respectively 10 and 100 times smaller than the errors of both sets of random controls. Similar results have been achieved for other smooth distributions, which show the efficiency of the two proposed algorithms. Figures~\ref{fig: double peak} and ~\ref{fig: double peak time} display respectively the true distribution and the minimal solution for all control sets for fixed and variable control times.
We observe that the solutions computed with controls generated by any algorithm match the true distribution. On the other hand, RCC completely fails,
showing a third peak in the middle, while RCCt can at least identify the two peaks of the distribution.

\subsection{The case of a step distribution.}\label{secnumstep}
\begin{figure}[h]
	\centering
	\includegraphics[width = 0.65\textwidth]{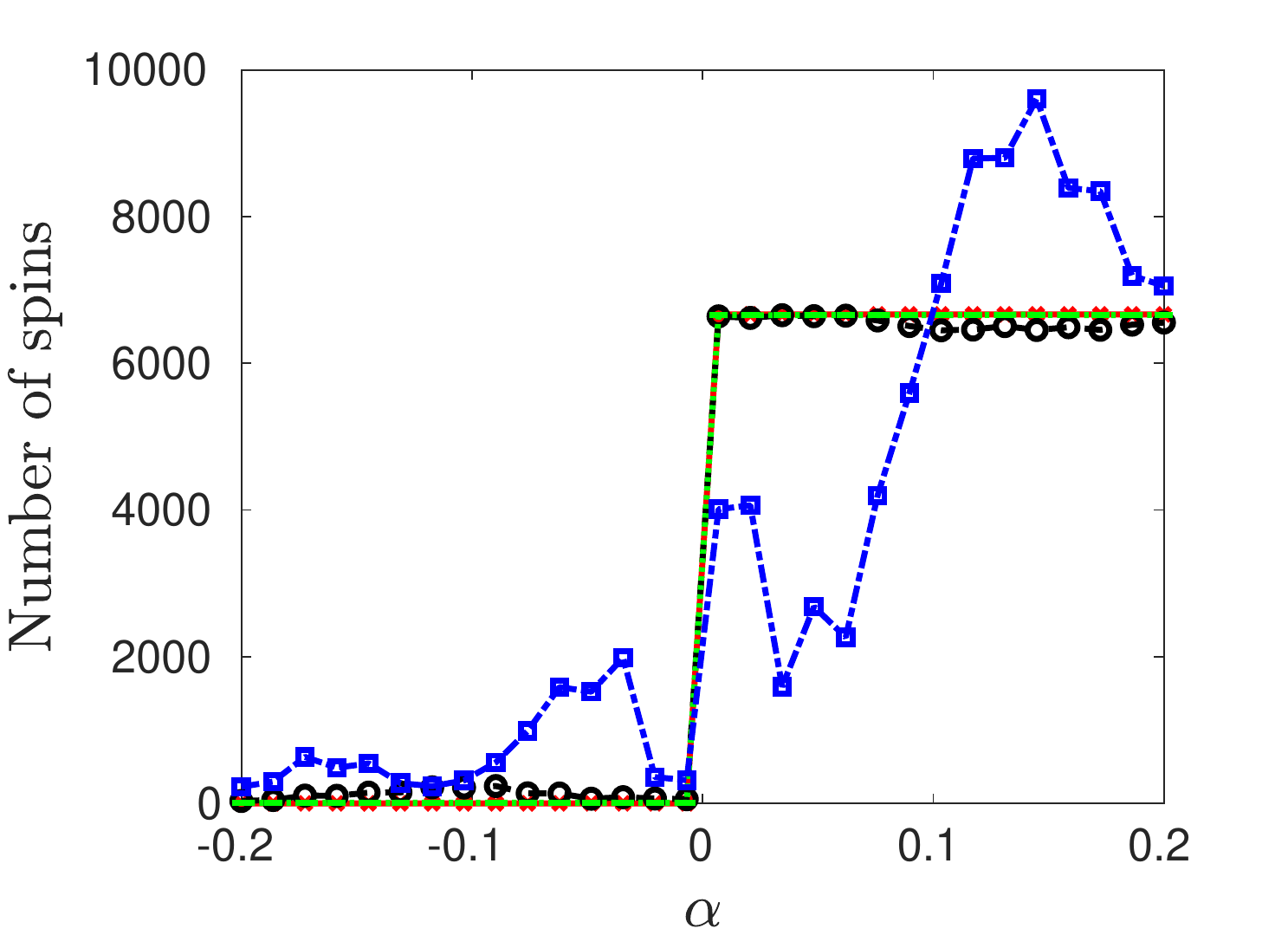}
	\caption{Same as Fig.~\ref{fig: double peak} but for a step distribution.}
	\label{fig: step distribution}
\end{figure}

\begin{figure}[h]
	\centering
	\includegraphics[width = 0.65\textwidth]{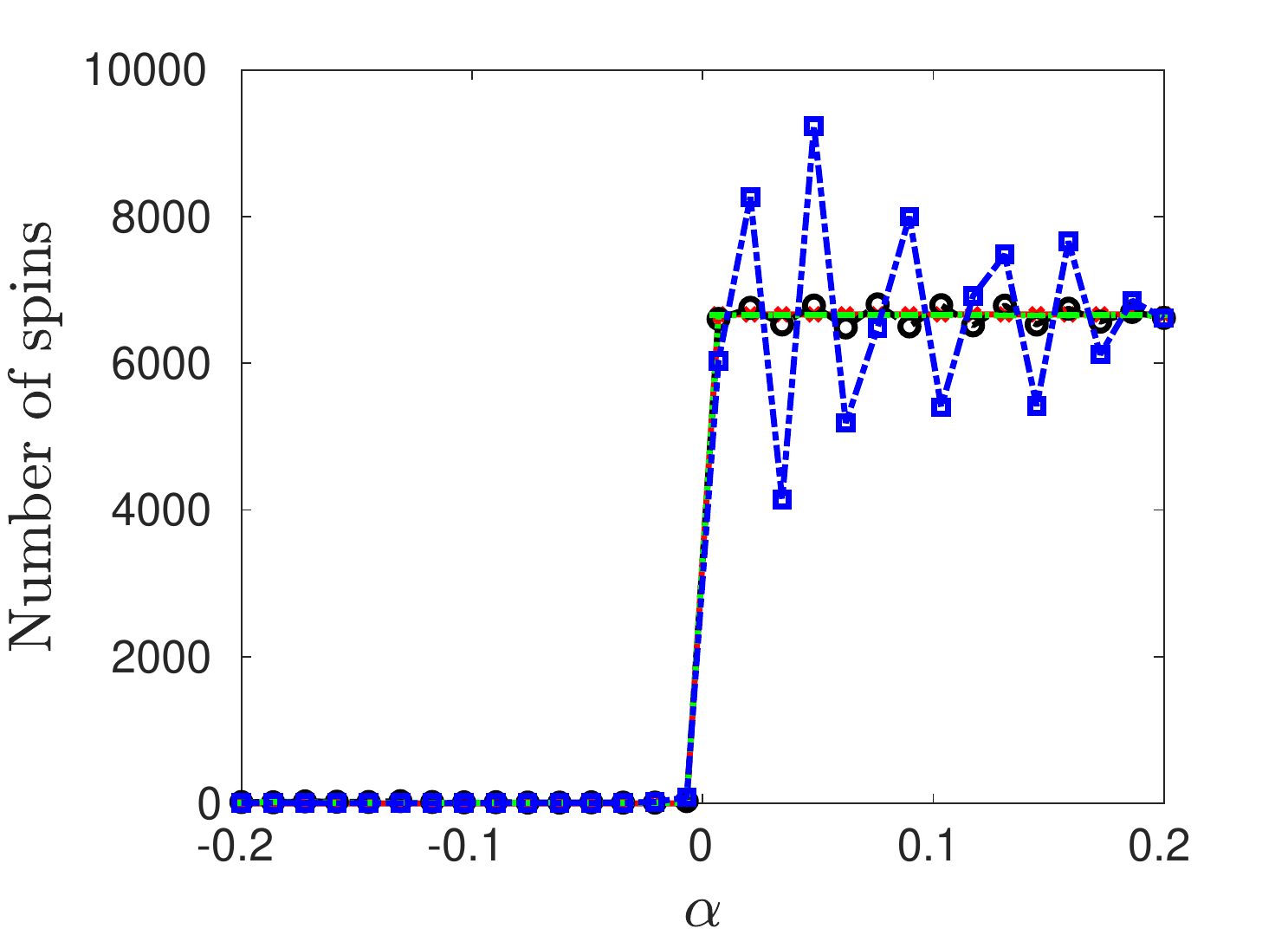}
	\caption{Same as Fig.~\ref{fig: step distribution}, but for variable control times.}
	\label{fig: step distribution time}
\end{figure}

As a second illustrative example, we consider a non-continuous step distribution, displayed in Fig.~\ref{fig: step distribution}, in which only spins with a positive parameter $\alpha$ can be observed in the sample.
We repeat the numerical simulations of Sec.~\ref{secnumgra} and we obtain the results reported in Tab.~\ref{tab:2}.
\begin{table}[h]
	\centering
	\begin{tabular}{ l|c|c|c|c|c|c }
		\hline
		Control set & GRA & GRAt & OGRA & OGRAt & RCC & RCCt \\ \hline\hline
		Min. error&0.0295&0.0181&0.0018&0.0021&0.4204&0.1943\\
		\hline
	\end{tabular}
	\caption{Minimum relative norm error for different control sets.}
	\label{tab:2}
\end{table}
As can be seen in Tab.~\ref{tab:2}, the difference in magnitude of errors is similar to the one for the double peak distribution. These results are displayed in Fig.~\ref{fig: step distribution} and ~\ref{fig: step distribution time}.

We observe that OGRA and OGRAt are still able to identify the true distribution, while GRA and GRAt already show small discrepancies. RCC completely fails again, but also RCCt shows major visible differences in the upper part of the step distribution. Arguments based on the properties of the matrix $W$ are given in Sec.~\ref{seceig} to explain such numerical observations.

\subsection{Eigenvalues and eigenvectors of $W$}\label{seceig}
\begin{figure}[h]
	\centering
	\includegraphics[width = 0.65\textwidth]{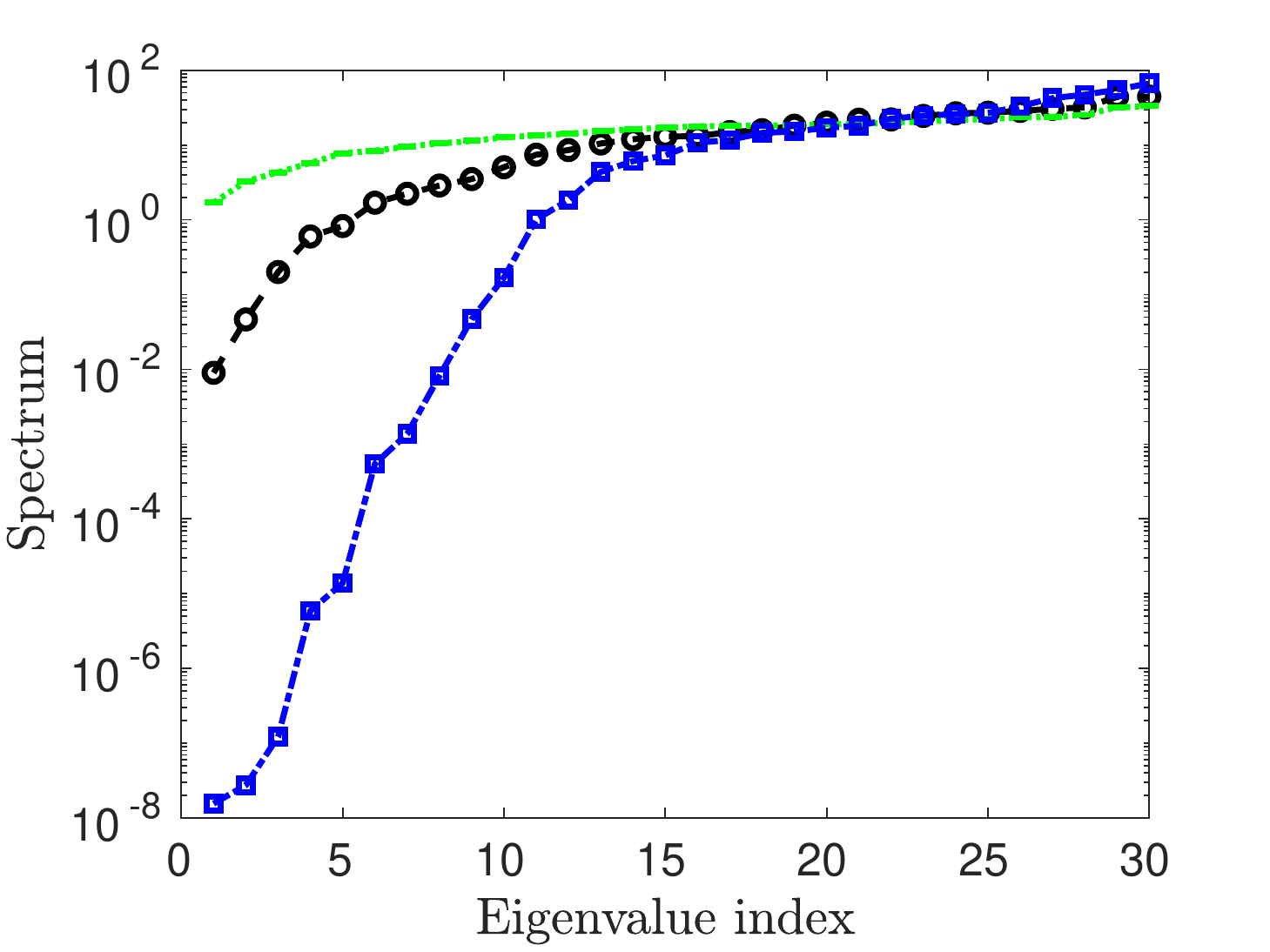}
	\caption{Spectrum of the matrix $W$ defined in \eqref{eq: W} for controls generated by OGRA (green, vertical lines), GRA (black, circles) and for RCC (blue, squares). Dimensionless units are used.}
	\label{fig: eigvals constant time}
\end{figure}

\begin{figure}[h]
	\centering
	\includegraphics[width = 0.65\textwidth]{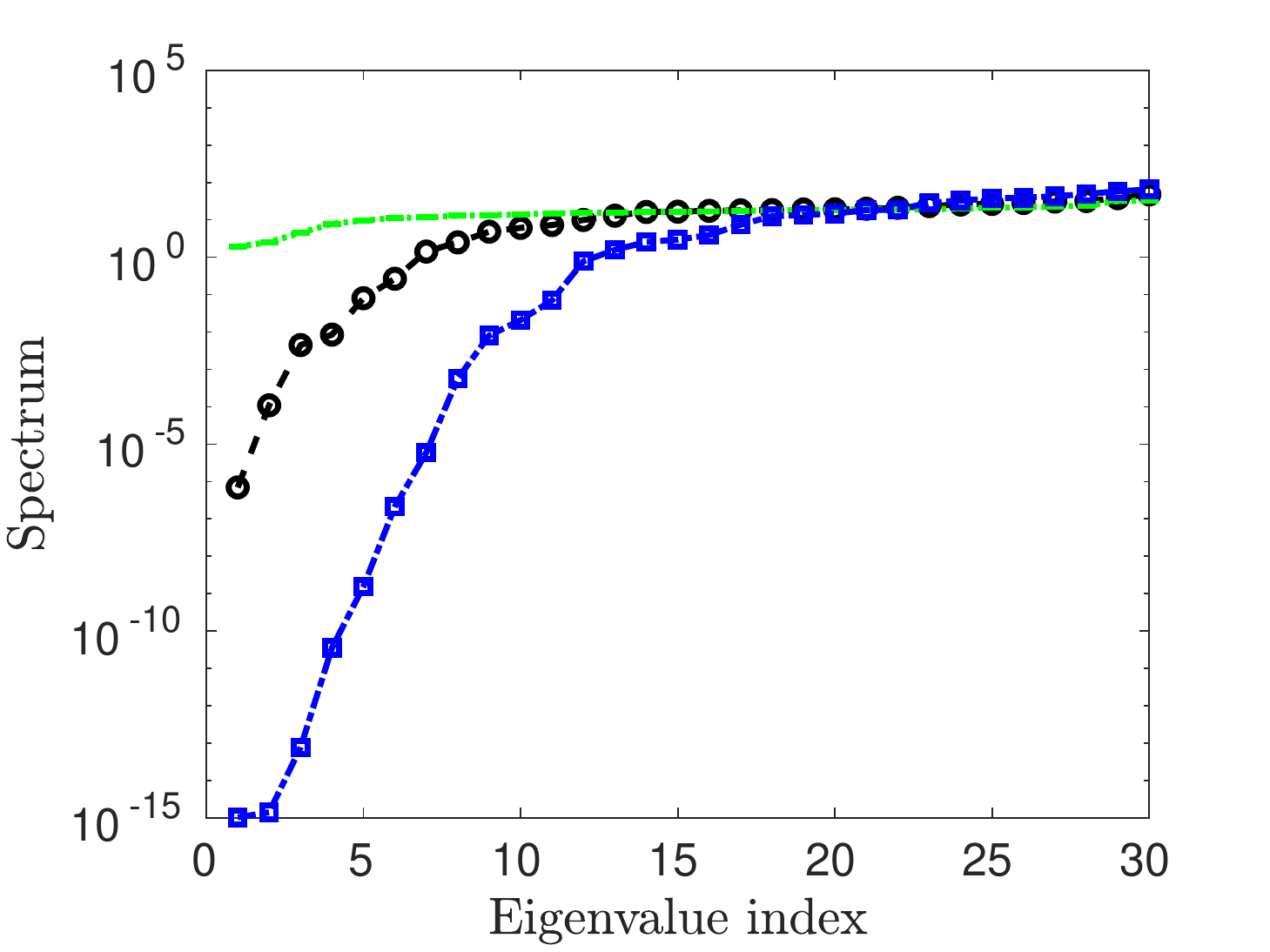}
	\caption{Same as Fig.~\ref{fig: eigvals constant time}, but for variable control times.}
	\label{fig: eigvals variable time}
\end{figure}

We explain qualitatively in this section the numerical results observed in Sec.~\ref{secnumgra} and \ref{secnumstep} through the properties of the matrix $W$, i.e. its eigenvalues and eigenvectors. We present the spectra of the matrix $W$ for different control sets in Fig.~\ref{fig: eigvals constant time} and~\ref{fig: eigvals variable time}.
A very large difference is observed between the eigenvalues associated with the optimized controls and the random ones. Note that this observation is the same if the control time is also optimized. This difference is quantitatively measured by the condition number of $W$, i.e. the ratio between the largest and the smallest eigenvalues, which is given in Tab~\ref{tab:4}. As could be expected, these results show that, using random controls, the matrix $W$ can be close to being singular. In the example of Fig.~\ref{fig: eigvals variable time}, while most of the eigenvalues are larger than 1, four of them are
smaller than $10^{-10}$. Hence, the matrix $W$ has a very bad condition number.
We stress the very good result achieved by OGRA for which all the eigenvalues have almost the same value. This analysis may also explain the difference between a smooth and a non-continuous probability distributions. As a matter of fact, numerical results reveal that random controls have more difficulty identifying non-smooth probability distribution as illustrated in Sec.~\ref{secnumstep}. This aspect can be understood from the behavior of the eigenvectors. Indeed, we observe numerically that the modes with a large number of oscillations correspond to the smallest eigvenvalues. Such modes have to be used to reconstruct probability distributions with rapid and abrupt variations. For random controls, these eigenvectors lead to large errors and to wrong probability distributions.


\begin{table}[h]
	\centering
	\begin{tabular}{ l|c|c|c|c|c|c }
		\hline
		Control set & GRA & GRAt & OGRA & OGRAt & RCC & RCCt \\ \hline\hline
		cond($W$)&4.9$\cdot10^3$&6.9$\cdot10^{7}$&19.55&16.6178&4.3$\cdot10^9$&1.42$\cdot10^{17}$\\
		\hline
	\end{tabular}
	\caption{Condition number of the matrix $W$ for different control sets.}
	\label{tab:4}
\end{table}

\section{Conclusions}\label{sec:conclusions}
We have introduced in this work a Greedy Reconstruction Algorithm with an application to spin dynamics. The algorithm provides
a systematic way to identify the probability distribution of a parameter of the Hamiltonian system varying in a given range. The efficiency of the identification process
has been illustrated in the case of a spin ensemble subjected to an inhomogeneous radio-frequency magnetic field. After having described some mathematical properties of the algorithm, numerical simulations have revealed the efficiency of GRA and its quite large basin of convergence. We have shown that GRA is able to identify non-trivial probability distributions with several peaks or with a step variation. An optimized version of this algorithm can be derived to further improve the identification process. We have limited the study to constant controls, but similar results can be achieved with time-dependent pulses. A quantitative comparison with random constant controls have highlighted the non-trivial recognition process realized by the algorithms. The numerical observations can also be partly explained by the computation of the eigenvalues and eigenvectors of the matrix $W$.

This analysis paves the way for further investigations in magnetic resonance. An interesting direction is the study of the sensitivity of the algorithm to experimental imperfections or to the presence of noise. It could be also used to identify probability distribution of other parameters, such as the resonance offset. These greedy algorithms
could also be transferred to other domains such as quantum optics and atomic and molecular physics. Finally, we hope that our method will be used in relevant experimental applications in magnetic resonance in a near future.\\ \\
\noindent\textbf{Acknowledgment}\\
The work of the first author was supported by the DFG via the collaborative research center SFB1432, Project-ID 425217212.
This research has been partially supported by the ANR project ``QUACO'' ANR-17-CE40-0007-01.  This project has received funding from the European Union Horizon 2020 research and innovation program under Marie-Sklodowska-Curie Grant No. 765267 (QUSCO).

\appendix
\section{Mathematical description of GRA}\label{mathdesc}
We give in this section some mathematical details about GRA. Straightforward computations show that the different steps of GRA can be expressed in matrix form as follows:
\begin{itemize}
\item The initialization problem \eqref{eq: initialization} is equivalent to:
$$
\max_{\textbf{u}\in\mathcal{U}}[W(\textbf{u})]_{1,1}
$$
\item The fitting-step problem \eqref{eq: fitting step} is equivalent to:
$$
\min_{\beta\in \mathbb{R}^k}\langle \beta|W^k_{[1:k,1:k]}|\beta\rangle -2\langle W^k_{[1:k,k+1]}|\beta \rangle
$$
where $W^k=\sum_{m=1}^kW(u_m)$. $W^k_{[1:k,1:k]}$ and $W^k_{[1:k,k+1]}$ denote respectively the $k\times k$ upper-left block of $W^{k}$ and a column vector containing the first $k$ components of the $k+1$-th column of $W^k$.
\item The discriminatory-step problem \eqref{eq: discriminatory step} is equivalent to:
$$
\max_{u\in\mathcal{U}}\langle \textbf{v}|[W(\textbf{u})]_{[1:k+1,1:k+1]}|\textbf{v}\rangle
$$
where $\textbf{v}=(\beta_k^\intercal,-1)^\intercal$.
\end{itemize}
The different iterations of GRA can then be described as follows. At iteration $k$, we assume that the sub-matrix $W^k_{[1:k,1:k]}$ is positive definite, but $W^k_{[1:k+1,1:k+1]}$ can have a non-trivial kernel. The idea is first to identify the kernel of $W^k_{[1:k+1,1:k+1]}$ by solving~\eqref{eq: fitting step} and then to compute a new control $\textbf{u}_{k+1}$ such that the new updated matrix $W^{k+1}=W^k+W(\textbf{u}_{k+1})$ has a positive definite upper-left block $W^{k+1}_{[1:k+1,1:k+1]}$. The convergence of the algorithm follows from this iterative process.

The following two technical lemmas describe the optimizations used in the two steps of the algorithm. In particular, Lemma \ref{lem1} shows that the fitting step identifies the kernel of the matrix $W^k_{[1:k+1,1:k+1]}$.
\begin{lemma}\label{lem1}
Assume that $W^k_{[1:k,1:k]}$ is positive definite and $W^k_{[1:k+1,1:k+1]}$ has a non-trivial kernel. Then the vector ${\bf v}=(\beta_k^\intercal,-1)^\intercal$, where $\beta_k$ is the solution to~\eqref{eq: fitting step}, is in the kernel of $W^k_{[1:k+1,1:k+1]}$.
\end{lemma}
The second lemma is the basis of the discriminatory-step algorithm and shows that this step corrects the rank deficiency of $W^k_{[1:k+1,1:k+1]}$.
\begin{lemma}
Let $W^k_{[1:k,1:k]}$ be a positive definite matrix and $\beta_k$ a solution of the fitting-step problem \eqref{eq: fitting step}. Any solution $\textbf{u}_{k+1}$ of \eqref{eq: discriminatory step} satisfies:
$$
\langle {\bf v}|W_{[1:k+1,1:k+1]}|{\bf v}\rangle >0
$$
for $k=0,1,\cdots, K-1$, where ${\bf v}=(\beta_k^\intercal,-1)^\intercal$.
\end{lemma}
The mathematical proofs of these results and a detailed numerical analysis of the GRA is beyond the scope of this work and will be presented elsewhere.

\section{The optimized greedy algorithm}\label{OGRA}
We discuss in this paragraph the optimized version of GRA.
It can been shown numerically that the behavior and the efficiency of GRA is strongly affected by the choice of the elements $\phi_k$ and their ordering.
GRA is essentially a sweep over the set $(\phi_k)_{k=1}^K$. However, a wrong choice of the elements $\phi_k$ and their ordering can lead to the stagnation
of the algorithm and to the computation of many useless control functions. Note that the stagnation of the algorithm can be measured in terms of rank corrections, i.e.
if for consecutive iterations the rank of $W$ does not increases. These reasons are at the origin of an optimized algorithm~\cite{BCS2021}.
OGRA takes as input a set $\Phi$, possibly larger than $(\phi_k)_{k=1}^K$ with linearly dependent elements, and returns as output not only
a set of $\widetilde{K}$ control functions, but also a set of linearly independent
functions $(\widehat{\phi}_k)_{k=1}^{\widehat{K}}$. The integers $\widetilde{K}$ and $\widehat{K}$ are
not necessarily equal and may be smaller than $K$ (in contrast to GRA).
The extension of the OGR method of \cite{BCS2021} to the distribution reconstruction problem
is detailed below, where we use the map $\textbf{h}_{\mathcal{S}}$ defined as
$$
\textbf{h}_{ \mathcal{S} } (\beta,\textbf{u}) = \sum_{\ell=1}^K \sum_{j=1}^{\textrm{card}[\mathcal{S}]} \beta_j \phi_j(\ell) \textbf{Y}(\textbf{u},\alpha_\ell),
$$
where $\mathcal{S} = ( \phi_1 ,\cdots , \phi_k )$. Note that for the fitting-step problem, we do not have any constraint for the choice of coefficients $\beta$.
This is due to the fact that, during the algorithm, we are not trying to reconstruct a distribution but to make the respective sub-matrix positive definite.\\ \\
\noindent {\bf Optimized Greedy Reconstruction Algorithm (OGRA):} Given a set of $K_+\geq K$ linearly independent functions $(\phi_1,\ldots,\phi_{K_+})$ and a tolerance $\textrm{tol}>0$.\\
Solve the initialization problem
\begin{equation}\label{eq: initialization:opt}
\max_{n\in\{1,\cdots,K_+\}}\max_{\textbf{u}\in\mathcal{U}} \|\textbf{h}_{\phi_n}^{(1)}(1,\textbf{u})\|^2,
\end{equation}
which gives the control $\textbf{u}_1$, and the control $\ell_1$. Set $k=1$ and $\mathcal{S} = \{ \phi_{\ell_1} \}$,
		$\widetilde{K} = K_+$, and update $\Phi = \Phi \setminus \{ \phi_{\ell_1} \}$. The algorithm is stopped if $\|\textbf{h}_{\mathcal{S}}^{(1)}(1,\textbf{u})\|^2<\textrm{tol}$.\\
\textbf{While $k\leq K-1$ do}
\begin{enumerate}
\item Remove elements from $\Phi$ that are linearly dependent on the ones in $\mathcal{S}$. Shift the indices of the
	remaining elements in $\Phi$. Update $\textrm{card}[\Phi]\rightarrow \widetilde{K}$.
\item \textbf{for $\ell=1,\cdots,\widetilde{K}$ do}\\
\underline{Fitting step}: Find $(\beta^{\ell}_j)_{j=1,\dots,k}$ that solve the problem
		\begin{equation}\label{eq: fitting step2}
		\min_{\beta \in\mathbb{R}^k}\sum_{m=1}^{k}\| \textbf{h}_{\phi_\ell}(1,\textbf{u}_m)-\textbf{h}_{\mathcal{S}}(\beta,\textbf{u}_m)\|^2,
		\end{equation}
\textbf{end for}
\item \underline{Discriminatory step}: Find $\textbf{u}_{k+1}$ and $\ell_{k+1}$ that solve the problem
		\begin{equation}\label{eq: discriminatory step2}
		\max_{\ell\in\{1,\cdots,K_+\}}\max_{\textbf{u}\in\mathcal{U}}\|\textbf{h}_{\phi_\ell}(1,\textbf{u})-\textbf{h}_{\mathcal{S}}(\beta^\ell,\textbf{u})\|^2.
		\end{equation}
If $\|\textbf{h}_{\phi_\ell}(1,\textbf{u}^{k+1})-\textbf{h}_{\mathcal{S}}(\beta^{\ell_{k+1}},\textbf{u}_{k+1})\|^2<\textrm{tol}$ then stop and return $\mathcal{S}$ and the computed $(\textbf{u})_{m=1}^k$.
\item Orthogonalize the function $\phi_{\ell_{k+1}}$ with respect to $\mathcal{S}$ and update $\mathcal{S} \cup  \{ \phi_{\ell_{k+1}} \}\rightarrow \mathcal{S}$, $\Phi \setminus \{ \phi_{\ell_{k+1}} \}\rightarrow \Phi$ and $k+1\rightarrow k$.
\end{enumerate}
\textbf{End while}\\ \\

\bibliographystyle{unsrt}

\begin{thebibliography}{61}%
\makeatletter
\providecommand \@ifxundefined [1]{%
 \@ifx{#1\undefined}
}%
\providecommand \@ifnum [1]{%
 \ifnum #1\expandafter \@firstoftwo
 \else \expandafter \@secondoftwo
 \fi
}%
\providecommand \@ifx [1]{%
 \ifx #1\expandafter \@firstoftwo
 \else \expandafter \@secondoftwo
 \fi
}%
\providecommand \natexlab [1]{#1}%
\providecommand \enquote  [1]{``#1''}%
\providecommand \bibnamefont  [1]{#1}%
\providecommand \bibfnamefont [1]{#1}%
\providecommand \citenamefont [1]{#1}%
\providecommand \href@noop [0]{\@secondoftwo}%
\providecommand \href [0]{\begingroup \@sanitize@url \@href}%
\providecommand \@href[1]{\@@startlink{#1}\@@href}%
\providecommand \@@href[1]{\endgroup#1\@@endlink}%
\providecommand \@sanitize@url [0]{\catcode `\\12\catcode `\$12\catcode
  `\&12\catcode `\#12\catcode `\^12\catcode `\_12\catcode `\%12\relax}%
\providecommand \@@startlink[1]{}%
\providecommand \@@endlink[0]{}%
\providecommand \url  [0]{\begingroup\@sanitize@url \@url }%
\providecommand \@url [1]{\endgroup\@href {#1}{\urlprefix }}%
\providecommand \urlprefix  [0]{URL }%
\providecommand \Eprint [0]{\href }%
\providecommand \doibase [0]{https://doi.org/}%
\providecommand \selectlanguage [0]{\@gobble}%
\providecommand \bibinfo  [0]{\@secondoftwo}%
\providecommand \bibfield  [0]{\@secondoftwo}%
\providecommand \translation [1]{[#1]}%
\providecommand \BibitemOpen [0]{}%
\providecommand \bibitemStop [0]{}%
\providecommand \bibitemNoStop [0]{.\EOS\space}%
\providecommand \EOS [0]{\spacefactor3000\relax}%
\providecommand \BibitemShut  [1]{\csname bibitem#1\endcsname}%
\let\auto@bib@innerbib\@empty
\bibitem{brif:2010}%
  \BibitemOpen
  \bibfield  {author} {\bibinfo {author} {\bibfnamefont {C.}~\bibnamefont
  {Brif}}, \bibinfo {author} {\bibfnamefont {R.}~\bibnamefont {Chakrabarti}},\
  and\ \bibinfo {author} {\bibfnamefont {R.}~\bibnamefont {Rabitz}},\
  }\href@noop {} {\bibfield  {journal} {\bibinfo  {journal} {New J. Phys.}\
  }\textbf {\bibinfo {volume} {12}},\ \bibinfo {pages} {075008} (\bibinfo
  {year} {2010})}\BibitemShut {NoStop}%
\bibitem{glaser_training_2015}%
  \BibitemOpen
  \bibfield  {author} {\bibinfo {author} {\bibfnamefont {S.~J.}\ \bibnamefont
  {Glaser}}, \bibinfo {author} {\bibfnamefont {U.}~\bibnamefont {Boscain}},
  \bibinfo {author} {\bibfnamefont {T.}~\bibnamefont {Calarco}}, \bibinfo
  {author} {\bibfnamefont {C.~P.}\ \bibnamefont {Koch}}, \bibinfo {author}
  {\bibfnamefont {W.}~\bibnamefont {Köckenberger}}, \bibinfo {author}
  {\bibfnamefont {R.}~\bibnamefont {Kosloff}}, \bibinfo {author} {\bibfnamefont
  {I.}~\bibnamefont {Kuprov}}, \bibinfo {author} {\bibfnamefont
  {B.}~\bibnamefont {Luy}}, \bibinfo {author} {\bibfnamefont {S.}~\bibnamefont
  {Schirmer}}, \bibinfo {author} {\bibfnamefont {T.}~\bibnamefont
  {Schulte-Herbrüggen}}, \bibinfo {author} {\bibfnamefont {D.}~\bibnamefont
  {Sugny}},\ and\ \bibinfo {author} {\bibfnamefont {F.~K.}\ \bibnamefont
  {Wilhelm}},\ } {\bibfield
  {journal} {\bibinfo  {journal} {Eur. Phys. J. D}\ }\textbf {\bibinfo {volume}
  {69}},\ \bibinfo {pages} {279} (\bibinfo {year} {2015})}\BibitemShut
  {NoStop}%
\bibitem{RMP:rotation}%
  \BibitemOpen
  \bibfield  {author} {\bibinfo {author} {\bibfnamefont {C.~P.}\ \bibnamefont
  {Koch}}, \bibinfo {author} {\bibfnamefont {M.}~\bibnamefont {Lemeshko}},\
  and\ \bibinfo {author} {\bibfnamefont {D.}~\bibnamefont {Sugny}},\ } {\bibfield  {journal}
  {\bibinfo  {journal} {Rev. Mod. Phys.}\ }\textbf {\bibinfo {volume} {91}},\
  \bibinfo {pages} {035005} (\bibinfo {year} {2019})}\BibitemShut {NoStop}%
\bibitem{koch_controlling_2016}%
  \BibitemOpen
  \bibfield  {author} {\bibinfo {author} {\bibfnamefont {C.~P.}\ \bibnamefont
  {Koch}},\ } {\bibfield
  {journal} {\bibinfo  {journal} {J. Phys. Condens. Matter}\ }\textbf {\bibinfo
  {volume} {28}},\ \bibinfo {pages} {213001} (\bibinfo {year}
  {2016})}\BibitemShut {NoStop}%
\bibitem{BCSbook}%
  \BibitemOpen
  \bibfield  {author} {\bibinfo {author} {\bibfnamefont {A.}~\bibnamefont
  {Borzì}}, \bibinfo {author} {\bibfnamefont {G.}~\bibnamefont {Ciaramella}},\
  and\ \bibinfo {author} {\bibfnamefont {M.}~\bibnamefont {Sprengel}},\
  }\href@noop {} {\emph {\bibinfo {title} {Formulation and Numerical Solution
  of Quantum Control Problems}}}\ (\bibinfo  {publisher} {SIAM},\ \bibinfo
  {address} {Philadelphia, PA},\ \bibinfo {year} {2017})\BibitemShut {NoStop}%
\bibitem{monschemes}%
  \BibitemOpen
  \bibfield  {author} {\bibinfo {author} {\bibfnamefont {Y.}~\bibnamefont
  {Maday}}, \bibinfo {author} {\bibfnamefont {J.}~\bibnamefont {Salomon}},\
  and\ \bibinfo {author} {\bibfnamefont {G.}~\bibnamefont {Turinici}},\
  }\href@noop {} {\bibfield  {journal} {\bibinfo  {journal} {Numer. Math.}\
  }\textbf {\bibinfo {volume} {103}},\ \bibinfo {pages} {323} (\bibinfo {year}
  {2006})}\BibitemShut {NoStop}%
\bibitem{sprengel2018investigation}%
  \BibitemOpen
  \bibfield  {author} {\bibinfo {author} {\bibfnamefont {M.}~\bibnamefont
  {Sprengel}}, \bibinfo {author} {\bibfnamefont {G.}~\bibnamefont
  {Ciaramella}},\ and\ \bibinfo {author} {\bibfnamefont {A.}~\bibnamefont
  {Borz{\`i}}},\ }\href@noop {} {\bibfield  {journal} {\bibinfo  {journal}
  {Journal of Dynamical and Control Systems}\ }\textbf {\bibinfo {volume}
  {24}},\ \bibinfo {pages} {657} (\bibinfo {year} {2018})}\BibitemShut
  {NoStop}%
\bibitem{BSS}%
  \BibitemOpen
  \bibfield  {author} {\bibinfo {author} {\bibfnamefont {U.}~\bibnamefont
  {Boscain}}, \bibinfo {author} {\bibfnamefont {M.}~\bibnamefont {Sigalotti}},\
  and\ \bibinfo {author} {\bibfnamefont {D.}~\bibnamefont {Sugny}},\
  }\href@noop {} {\bibfield  {journal} {\bibinfo  {journal} {to be published in
  PRX Quantum}\ } (\bibinfo {year} {2021})}\BibitemShut {NoStop}%
\bibitem{QT}%
  \BibitemOpen
  \bibfield  {author} {\bibinfo {author} {\bibfnamefont {A.}~\bibnamefont
  {Ac{\'{\i}}n}}, \bibinfo {author} {\bibfnamefont {I.}~\bibnamefont {Bloch}},
  \bibinfo {author} {\bibfnamefont {H.}~\bibnamefont {Buhrman}}, \bibinfo
  {author} {\bibfnamefont {T.}~\bibnamefont {Calarco}}, \bibinfo {author}
  {\bibfnamefont {C.}~\bibnamefont {Eichler}}, \bibinfo {author} {\bibfnamefont
  {J.}~\bibnamefont {Eisert}}, \bibinfo {author} {\bibfnamefont
  {D.}~\bibnamefont {Esteve}}, \bibinfo {author} {\bibfnamefont
  {N.}~\bibnamefont {Gisin}}, \bibinfo {author} {\bibfnamefont {S.~J.}\
  \bibnamefont {Glaser}}, \bibinfo {author} {\bibfnamefont {F.}~\bibnamefont
  {Jelezko}}, \bibinfo {author} {\bibfnamefont {S.}~\bibnamefont {Kuhr}},
  \bibinfo {author} {\bibfnamefont {M.}~\bibnamefont {Lewenstein}}, \bibinfo
  {author} {\bibfnamefont {M.~F.}\ \bibnamefont {Riedel}}, \bibinfo {author}
  {\bibfnamefont {P.~O.}\ \bibnamefont {Schmidt}}, \bibinfo {author}
  {\bibfnamefont {R.}~\bibnamefont {Thew}}, \bibinfo {author} {\bibfnamefont
  {A.}~\bibnamefont {Wallraff}}, \bibinfo {author} {\bibfnamefont
  {I.}~\bibnamefont {Walmsley}},\ and\ \bibinfo {author} {\bibfnamefont
  {F.~K.}\ \bibnamefont {Wilhelm}},\ } {\bibfield  {journal} {\bibinfo
  {journal} {New J. Phys.}\ }\textbf {\bibinfo {volume} {20}},\ \bibinfo
  {pages} {080201} (\bibinfo {year} {2018})}\BibitemShut {NoStop}%
\bibitem{alessandrobook}%
  \BibitemOpen
  \bibfield  {author} {\bibinfo {author} {\bibfnamefont {D.}~\bibnamefont
  {D'Alessandro}},\ }\href@noop {} {\emph {\bibinfo {title} {Introduction to
  Quantum Control and Dynamics}}}\ (\bibinfo  {publisher} {Chapman \& Hall/CRC,
  Boca Raton},\ \bibinfo {year} {2007})\BibitemShut {NoStop}%
\bibitem{brysonbook}%
  \BibitemOpen
  \bibfield  {author} {\bibinfo {author} {\bibfnamefont {A.~E.}\ \bibnamefont
  {Bryson}}\ and\ \bibinfo {author} {\bibfnamefont {Y.-C.}\ \bibnamefont
  {Ho}},\ }\href@noop {} {\emph {\bibinfo {title} {Applied Optimal Control:
  Optimization, Estimation, and Control}}}\ (\bibinfo  {publisher} {Hemisphere,
  Washington, DC},\ \bibinfo {year} {1975})\BibitemShut {NoStop}%
\bibitem{geremia2003}%
  \BibitemOpen
  \bibfield  {author} {\bibinfo {author} {\bibfnamefont {J.~M.}\ \bibnamefont
  {Geremia}}\ and\ \bibinfo {author} {\bibfnamefont {H.}~\bibnamefont
  {Rabitz}},\ } {\bibfield  {journal}
  {\bibinfo  {journal} {The Journal of Chemical Physics}\ }\textbf {\bibinfo
  {volume} {118}},\ \bibinfo {pages} {5369} (\bibinfo {year}
  {2003})}\BibitemShut {NoStop}%
\bibitem{feng2006}%
  \BibitemOpen
  \bibfield  {author} {\bibinfo {author} {\bibfnamefont {A.}~\bibnamefont
  {Feng}}, \bibinfo {author} {\bibfnamefont {H.}~\bibnamefont {Rabitz}},
  \bibinfo {author} {\bibfnamefont {G.}~\bibnamefont {Turinici}},\ and\
  \bibinfo {author} {\bibfnamefont {C.}~\bibnamefont {Le~Bris}},\ }\href@noop
  {} {\bibfield  {journal} {\bibinfo  {journal} {J. Phys. Chem. A}\ }\textbf
  {\bibinfo {volume} {110}},\ \bibinfo {pages} {7755} (\bibinfo {year}
  {2006})}\BibitemShut {NoStop}%
\bibitem{geremia2002}%
  \BibitemOpen
  \bibfield  {author} {\bibinfo {author} {\bibfnamefont {J.~M.}\ \bibnamefont
  {Geremia}}\ and\ \bibinfo {author} {\bibfnamefont {H.}~\bibnamefont
  {Rabitz}},\ } {\bibfield
   {journal} {\bibinfo  {journal} {Phys. Rev. Lett.}\ }\textbf {\bibinfo
  {volume} {89}},\ \bibinfo {pages} {263902} (\bibinfo {year}
  {2002})}\BibitemShut {NoStop}%
\bibitem{ma2013}%
  \BibitemOpen
  \bibfield  {author} {\bibinfo {author} {\bibfnamefont {D.}~\bibnamefont
  {Ma}}, \bibinfo {author} {\bibfnamefont {V.}~\bibnamefont {Gulani}},\ and\
  \bibinfo {author} {\bibfnamefont {N.}~\bibnamefont {Seiberlich}},\
  }\href@noop {} {\bibfield  {journal} {\bibinfo  {journal} {Nature}\ }\textbf
  {\bibinfo {volume} {495}},\ \bibinfo {pages} {187} (\bibinfo {year}
  {2013})}\BibitemShut {NoStop}%
\bibitem{ansel2017}%
  \BibitemOpen
  \bibfield  {author} {\bibinfo {author} {\bibfnamefont {Q.}~\bibnamefont
  {Ansel}}, \bibinfo {author} {\bibfnamefont {M.}~\bibnamefont {Tesch}},
  \bibinfo {author} {\bibfnamefont {S.~J.}\ \bibnamefont {Glaser}},\ and\
  \bibinfo {author} {\bibfnamefont {D.}~\bibnamefont {Sugny}},\ } {\bibfield  {journal} {\bibinfo
  {journal} {Phys. Rev. A}\ }\textbf {\bibinfo {volume} {96}},\ \bibinfo
  {pages} {053419} (\bibinfo {year} {2017})}\BibitemShut {NoStop}%
\bibitem{pierre2016}%
  \BibitemOpen
  \bibfield  {author} {\bibinfo {author} {\bibfnamefont {E.~Y.}\ \bibnamefont
  {Pierre}}, \bibinfo {author} {\bibfnamefont {D.}~\bibnamefont {Ma}}, \bibinfo
  {author} {\bibfnamefont {Y.}~\bibnamefont {Chen}}, \bibinfo {author}
  {\bibfnamefont {C.}~\bibnamefont {Badve}},\ and\ \bibinfo {author}
  {\bibfnamefont {M.~A.}\ \bibnamefont {Griswold}},\ } {\bibfield  {journal}
  {\bibinfo  {journal} {Magnetic Resonance in Medicine}\ }\textbf {\bibinfo
  {volume} {75}},\ \bibinfo {pages} {2481} (\bibinfo {year}
  {2016})}\BibitemShut {NoStop}%
\bibitem{CBDW2015}%
  \BibitemOpen
  \bibfield  {author} {\bibinfo {author} {\bibfnamefont {G.}~\bibnamefont
  {Ciaramella}}, \bibinfo {author} {\bibfnamefont {A.}~\bibnamefont {Borzì}},
  \bibinfo {author} {\bibfnamefont {G.}~\bibnamefont {Dirr}},\ and\ \bibinfo
  {author} {\bibfnamefont {D.}~\bibnamefont {Wachsmuth}},\ } {\bibfield  {journal} {\bibinfo
  {journal} {SIAM Journal on Scientific Computing}\ }\textbf {\bibinfo {volume}
  {37}},\ \bibinfo {pages} {A319} (\bibinfo {year} {2015})}\BibitemShut
  {NoStop}%
\bibitem{schirmer2004}%
  \BibitemOpen
  \bibfield  {author} {\bibinfo {author} {\bibfnamefont {S.~G.}\ \bibnamefont
  {Schirmer}}, \bibinfo {author} {\bibfnamefont {A.}~\bibnamefont {Kolli}},\
  and\ \bibinfo {author} {\bibfnamefont {D.~K.~L.}\ \bibnamefont {Oi}},\ } {\bibfield  {journal} {\bibinfo
  {journal} {Phys. Rev. A}\ }\textbf {\bibinfo {volume} {69}},\ \bibinfo
  {pages} {050306} (\bibinfo {year} {2004})}\BibitemShut {NoStop}%
\bibitem{schirmer2009}%
  \BibitemOpen
  \bibfield  {author} {\bibinfo {author} {\bibfnamefont {S.~G.}\ \bibnamefont
  {Schirmer}}\ and\ \bibinfo {author} {\bibfnamefont {D.~K.~L.}\ \bibnamefont
  {Oi}},\ } {\bibfield
  {journal} {\bibinfo  {journal} {Phys. Rev. A}\ }\textbf {\bibinfo {volume}
  {80}},\ \bibinfo {pages} {022333} (\bibinfo {year} {2009})}\BibitemShut
  {NoStop}%
\bibitem{schirmer2015}%
  \BibitemOpen
  \bibfield  {author} {\bibinfo {author} {\bibfnamefont {S.~G.}\ \bibnamefont
  {Schirmer}}\ and\ \bibinfo {author} {\bibfnamefont {F.~C.}\ \bibnamefont
  {Langbein}},\ } {\bibfield
  {journal} {\bibinfo  {journal} {Phys. Rev. A}\ }\textbf {\bibinfo {volume}
  {91}},\ \bibinfo {pages} {022125} (\bibinfo {year} {2015})}\BibitemShut
  {NoStop}%
\bibitem{yuan2017}%
  \BibitemOpen
  \bibfield  {author} {\bibinfo {author} {\bibfnamefont {H.}~\bibnamefont
  {Yuan}}\ and\ \bibinfo {author} {\bibfnamefont {C.~F.}\ \bibnamefont
  {Fung}},\ }\href@noop {} {\bibfield  {journal} {\bibinfo  {journal} {npj
  Quantum Inf.}\ }\textbf {\bibinfo {volume} {3}},\ \bibinfo {pages} {14}
  (\bibinfo {year} {2017})}\BibitemShut {NoStop}%
\bibitem{liu2017}%
  \BibitemOpen
  \bibfield  {author} {\bibinfo {author} {\bibfnamefont {J.}~\bibnamefont
  {Liu}}\ and\ \bibinfo {author} {\bibfnamefont {H.}~\bibnamefont {Yuan}},\
  } {\bibfield  {journal}
  {\bibinfo  {journal} {Phys. Rev. A}\ }\textbf {\bibinfo {volume} {96}},\
  \bibinfo {pages} {012117} (\bibinfo {year} {2017})}\BibitemShut {NoStop}%
\bibitem{wittler2021}%
  \BibitemOpen
  \bibfield  {author} {\bibinfo {author} {\bibfnamefont {N.}~\bibnamefont
  {Wittler}}, \bibinfo {author} {\bibfnamefont {F.}~\bibnamefont {Roy}},
  \bibinfo {author} {\bibfnamefont {K.}~\bibnamefont {Pack}}, \bibinfo {author}
  {\bibfnamefont {M.}~\bibnamefont {Werninghaus}}, \bibinfo {author}
  {\bibfnamefont {A.~S.}\ \bibnamefont {Roy}}, \bibinfo {author} {\bibfnamefont
  {D.~J.}\ \bibnamefont {Egger}}, \bibinfo {author} {\bibfnamefont
  {S.}~\bibnamefont {Filipp}}, \bibinfo {author} {\bibfnamefont {F.~K.}\
  \bibnamefont {Wilhelm}},\ and\ \bibinfo {author} {\bibfnamefont
  {S.}~\bibnamefont {Machnes}},\ } {\bibfield  {journal}
  {\bibinfo  {journal} {Phys. Rev. Applied}\ }\textbf {\bibinfo {volume}
  {15}},\ \bibinfo {pages} {034080} (\bibinfo {year} {2021})}\BibitemShut
  {NoStop}%
\bibitem{zhang2014}%
  \BibitemOpen
  \bibfield  {author} {\bibinfo {author} {\bibfnamefont {J.}~\bibnamefont
  {Zhang}}\ and\ \bibinfo {author} {\bibfnamefont {M.}~\bibnamefont
  {Sarovar}},\ } {\bibfield  {journal} {\bibinfo  {journal} {Phys. Rev. Lett.}\ }\textbf
  {\bibinfo {volume} {113}},\ \bibinfo {pages} {080401} (\bibinfo {year}
  {2014})}\BibitemShut {NoStop}%
\bibitem{sone2017}%
  \BibitemOpen
  \bibfield  {author} {\bibinfo {author} {\bibfnamefont {A.}~\bibnamefont
  {Sone}}\ and\ \bibinfo {author} {\bibfnamefont {P.}~\bibnamefont
  {Cappellaro}},\ } {\bibfield  {journal} {\bibinfo  {journal} {Phys. Rev. A}\ }\textbf {\bibinfo
  {volume} {95}},\ \bibinfo {pages} {022335} (\bibinfo {year}
  {2017})}\BibitemShut {NoStop}%
\bibitem{kiukas2017}%
  \BibitemOpen
  \bibfield  {author} {\bibinfo {author} {\bibfnamefont {J.}~\bibnamefont
  {Kiukas}}, \bibinfo {author} {\bibfnamefont {K.}~\bibnamefont {Yuasa}},\ and\
  \bibinfo {author} {\bibfnamefont {D.}~\bibnamefont {Burgarth}},\ } {\bibfield  {journal} {\bibinfo
  {journal} {Phys. Rev. A}\ }\textbf {\bibinfo {volume} {95}},\ \bibinfo
  {pages} {052132} (\bibinfo {year} {2017})}\BibitemShut {NoStop}%
\bibitem{burgarth2017}%
  \BibitemOpen
  \bibfield  {author} {\bibinfo {author} {\bibfnamefont {D.}~\bibnamefont
  {Burgarth}}\ and\ \bibinfo {author} {\bibfnamefont {A.}~\bibnamefont
  {Ajoy}},\ } {\bibfield
  {journal} {\bibinfo  {journal} {Phys. Rev. Lett.}\ }\textbf {\bibinfo
  {volume} {119}},\ \bibinfo {pages} {030402} (\bibinfo {year}
  {2017})}\BibitemShut {NoStop}%
\bibitem{Wang_2015}%
  \BibitemOpen
  \bibfield  {author} {\bibinfo {author} {\bibfnamefont {S.-T.}\ \bibnamefont
  {Wang}}, \bibinfo {author} {\bibfnamefont {D.-L.}\ \bibnamefont {Deng}},\
  and\ \bibinfo {author} {\bibfnamefont {L.-M.}\ \bibnamefont {Duan}},\ } {\bibfield  {journal}
  {\bibinfo  {journal} {New Journal of Physics}\ }\textbf {\bibinfo {volume}
  {17}},\ \bibinfo {pages} {093017} (\bibinfo {year} {2015})}\BibitemShut
  {NoStop}%
\bibitem{xue2021}%
  \BibitemOpen
  \bibfield  {author} {\bibinfo {author} {\bibfnamefont {S.}~\bibnamefont
  {Xue}}, \bibinfo {author} {\bibfnamefont {R.}~\bibnamefont {Wu}}, \bibinfo
  {author} {\bibfnamefont {S.}~\bibnamefont {Ma}}, \bibinfo {author}
  {\bibfnamefont {D.}~\bibnamefont {Li}},\ and\ \bibinfo {author}
  {\bibfnamefont {M.}~\bibnamefont {Jiang}},\ } {\bibfield  {journal} {\bibinfo
   {journal} {Phys. Rev. A}\ }\textbf {\bibinfo {volume} {103}},\ \bibinfo
  {pages} {022604} (\bibinfo {year} {2021})}\BibitemShut {NoStop}%
\bibitem{xue2020}%
  \BibitemOpen
  \bibfield  {author} {\bibinfo {author} {\bibfnamefont {S.}~\bibnamefont
  {Xue}}, \bibinfo {author} {\bibfnamefont {L.}~\bibnamefont {Tan}}, \bibinfo
  {author} {\bibfnamefont {R.}~\bibnamefont {Wu}}, \bibinfo {author}
  {\bibfnamefont {M.}~\bibnamefont {Jiang}},\ and\ \bibinfo {author}
  {\bibfnamefont {I.~R.}\ \bibnamefont {Petersen}},\ } {\bibfield  {journal} {\bibinfo
   {journal} {Phys. Rev. A}\ }\textbf {\bibinfo {volume} {102}},\ \bibinfo
  {pages} {042227} (\bibinfo {year} {2020})}\BibitemShut {NoStop}%
\bibitem{zhang2015}%
  \BibitemOpen
  \bibfield  {author} {\bibinfo {author} {\bibfnamefont {J.}~\bibnamefont
  {Zhang}}\ and\ \bibinfo {author} {\bibfnamefont {M.}~\bibnamefont
  {Sarovar}},\ } {\bibfield
  {journal} {\bibinfo  {journal} {Phys. Rev. A}\ }\textbf {\bibinfo {volume}
  {91}},\ \bibinfo {pages} {052121} (\bibinfo {year} {2015})}\BibitemShut
  {NoStop}%
\bibitem{Bennink_2019}%
  \BibitemOpen
  \bibfield  {author} {\bibinfo {author} {\bibfnamefont {R.~S.}\ \bibnamefont
  {Bennink}}\ and\ \bibinfo {author} {\bibfnamefont {P.}~\bibnamefont
  {Lougovski}},\ } {\bibfield
  {journal} {\bibinfo  {journal} {New Journal of Physics}\ }\textbf {\bibinfo
  {volume} {21}},\ \bibinfo {pages} {083013} (\bibinfo {year}
  {2019})}\BibitemShut {NoStop}%
\bibitem{ljung2010}%
  \BibitemOpen
  \bibfield  {author} {\bibinfo {author} {\bibfnamefont {L.}~\bibnamefont
  {Ljung}},\ } {\bibfield
  {journal} {\bibinfo  {journal} {Annual Reviews in Control}\ }\textbf
  {\bibinfo {volume} {34}},\ \bibinfo {pages} {1} (\bibinfo {year}
  {2010})}\BibitemShut {NoStop}%
\bibitem{lebris06}%
  \BibitemOpen
  \bibfield  {author} {\bibinfo {author} {\bibnamefont {{Le Bris, Claude}}},
  \bibinfo {author} {\bibnamefont {{Mirrahimi, Mazyar}}}, \bibinfo {author}
  {\bibnamefont {{Rabitz, Herschel}}},\ and\ \bibinfo {author} {\bibnamefont
  {{Turinici, Gabriel}}},\ }  {\bibfield  {journal} {\bibinfo  {journal} {ESAIM: COCV}\ }\textbf {\bibinfo
  {volume} {13}},\ \bibinfo {pages} {378} (\bibinfo {year} {2007})}\BibitemShut
  {NoStop}%
\bibitem{alis2004}%
  \BibitemOpen
  \bibfield  {author} {\bibinfo {author} {\bibfnamefont {O.~F.}\ \bibnamefont
  {Alis}}, \bibinfo {author} {\bibfnamefont {H.}~\bibnamefont {Rabiz}},
  \bibinfo {author} {\bibfnamefont {M.~Q.}\ \bibnamefont {Rosenthal}}, \bibinfo
  {author} {\bibfnamefont {C.}~\bibnamefont {Phan}},\ and\ \bibinfo {author}
  {\bibfnamefont {M.}~\bibnamefont {Pence}},\ }\href@noop {} {\bibfield
  {journal} {\bibinfo  {journal} {J. Math. Chem.}\ }\textbf {\bibinfo {volume}
  {35}},\ \bibinfo {pages} {65} (\bibinfo {year} {2004})}\BibitemShut {NoStop}%
\bibitem{ndong2014}%
  \BibitemOpen
  \bibfield  {author} {\bibinfo {author} {\bibfnamefont {M.}~\bibnamefont
  {Ndong}}, \bibinfo {author} {\bibfnamefont {J.}~\bibnamefont {Salomon}},\
  and\ \bibinfo {author} {\bibfnamefont {D.}~\bibnamefont {Sugny}},\
  }\href@noop {} {\bibfield  {journal} {\bibinfo  {journal} {J. Phys. A: Math.
  Theor.}\ }\textbf {\bibinfo {volume} {47}},\ \bibinfo {pages} {265302}
  (\bibinfo {year} {2014})}\BibitemShut {NoStop}%
\bibitem{rojas2007}%
  \BibitemOpen
  \bibfield  {author} {\bibinfo {author} {\bibfnamefont {C.~R.}\ \bibnamefont
  {Rojas}}, \bibinfo {author} {\bibfnamefont {J.~S.}\ \bibnamefont {Welsh}},
  \bibinfo {author} {\bibfnamefont {G.~C.}\ \bibnamefont {Goodwin}},\ and\
  \bibinfo {author} {\bibfnamefont {A.}~\bibnamefont {Feuer}},\ }{\bibfield
   {journal} {\bibinfo  {journal} {Automatica}\ }\textbf {\bibinfo {volume}
  {43}},\ \bibinfo {pages} {993} (\bibinfo {year} {2007})}\BibitemShut
  {NoStop}%
\bibitem{baudouin2008}%
  \BibitemOpen
  \bibfield  {author} {\bibinfo {author} {\bibfnamefont {L.}~\bibnamefont
  {Baudouin}}\ and\ \bibinfo {author} {\bibfnamefont {A.}~\bibnamefont
  {Mercado}},\ }\href@noop {} {\bibfield  {journal} {\bibinfo  {journal} {Appl.
  Anal.}\ }\textbf {\bibinfo {volume} {87}},\ \bibinfo {pages} {1145} (\bibinfo
  {year} {2008})}\BibitemShut {NoStop}%
\bibitem{bonnabel2009}%
  \BibitemOpen
  \bibfield  {author} {\bibinfo {author} {\bibfnamefont {S.}~\bibnamefont
  {Bonnabel}}, \bibinfo {author} {\bibfnamefont {M.}~\bibnamefont
  {Mirrahimi}},\ and\ \bibinfo {author} {\bibfnamefont {P.}~\bibnamefont
  {Rouchon}},\ }\href@noop {} {\bibfield  {journal} {\bibinfo  {journal}
  {Automatica}\ }\textbf {\bibinfo {volume} {45}},\ \bibinfo {pages} {1144}
  (\bibinfo {year} {2009})}\BibitemShut {NoStop}%
\bibitem{fu2017}%
  \BibitemOpen
  \bibfield  {author} {\bibinfo {author} {\bibfnamefont {Y.}~\bibnamefont
  {Fu}}\ and\ \bibinfo {author} {\bibfnamefont {G.}~\bibnamefont {Turinici}},\
  }\href@noop {} {\bibfield  {journal} {\bibinfo  {journal} {Appl. Anal.}\
  }\textbf {\bibinfo {volume} {23}},\ \bibinfo {pages} {1129} (\bibinfo {year}
  {2017})}\BibitemShut {NoStop}%
\bibitem{wang2018}%
  \BibitemOpen
  \bibfield  {author} {\bibinfo {author} {\bibfnamefont {Y.}~\bibnamefont
  {Wang}}, \bibinfo {author} {\bibfnamefont {D.}~\bibnamefont {Dong}}, \bibinfo
  {author} {\bibfnamefont {B.}~\bibnamefont {Qi}}, \bibinfo {author}
  {\bibfnamefont {J.}~\bibnamefont {Zhang}}, \bibinfo {author} {\bibfnamefont
  {I.~R.}\ \bibnamefont {Petersen}},\ and\ \bibinfo {author} {\bibfnamefont
  {H.}~\bibnamefont {Yonezawa}},\ }\href@noop {} {\bibfield  {journal}
  {\bibinfo  {journal} {IEEE Trans. Autom. Control}\ }\textbf {\bibinfo
  {volume} {63}},\ \bibinfo {pages} {1388} (\bibinfo {year}
  {2018})}\BibitemShut {NoStop}%
\bibitem{Li06}%
  \BibitemOpen
  \bibfield  {author} {\bibinfo {author} {\bibfnamefont {J.}~\bibnamefont
  {Li}}\ and\ \bibinfo {author} {\bibfnamefont {N.}~\bibnamefont {Khaneja}},\
  }\href@noop {} {\bibfield  {journal} {\bibinfo  {journal} {Phys. Rev. A}\
  }\textbf {\bibinfo {volume} {73}},\ \bibinfo {pages} {030302} (\bibinfo
  {year} {2006})}\BibitemShut {NoStop}%
\bibitem{li09}%
  \BibitemOpen
  \bibfield  {author} {\bibinfo {author} {\bibfnamefont {J.~S.}\ \bibnamefont
  {Li}}\ and\ \bibinfo {author} {\bibfnamefont {N.}~\bibnamefont {Khaneja}},\
  }\href@noop {} {\bibfield  {journal} {\bibinfo  {journal} {IEEE Trans. Auto.
  Control}\ }\textbf {\bibinfo {volume} {54}},\ \bibinfo {pages} {528}
  (\bibinfo {year} {2009})}\BibitemShut {NoStop}%
\bibitem{kobzar:2008}%
  \BibitemOpen
  \bibfield  {author} {\bibinfo {author} {\bibfnamefont {K.}~\bibnamefont
  {Kobzar}}, \bibinfo {author} {\bibfnamefont {T.~E.}\ \bibnamefont {Skinner}},
  \bibinfo {author} {\bibfnamefont {N.}~\bibnamefont {Khaneja}}, \bibinfo
  {author} {\bibfnamefont {S.~J.}\ \bibnamefont {Glaser}},\ and\ \bibinfo
  {author} {\bibfnamefont {B.}~\bibnamefont {Luy}},\ } {\bibfield
  {journal} {\bibinfo  {journal} {J. Magn. Reson.}\ }\textbf {\bibinfo {volume}
  {194}},\ \bibinfo {pages} {58} (\bibinfo {year} {2008})}\BibitemShut
  {NoStop}%
\bibitem{lapert_exploring_2012}%
  \BibitemOpen
  \bibfield  {author} {\bibinfo {author} {\bibfnamefont {M.}~\bibnamefont
  {Lapert}}, \bibinfo {author} {\bibfnamefont {Y.}~\bibnamefont {Zhang}},
  \bibinfo {author} {\bibfnamefont {M.~A.}\ \bibnamefont {Janich}}, \bibinfo
  {author} {\bibfnamefont {S.~J.}\ \bibnamefont {Glaser}},\ and\ \bibinfo
  {author} {\bibfnamefont {D.}~\bibnamefont {Sugny}},\ } {\bibinfo
  {journal} {Sci. Rep.}\ }\textbf {\bibinfo {volume} {2}},\ \bibinfo {pages}
  {589} (\bibinfo {year} {2012})\BibitemShut {NoStop}%
\bibitem{turinici2019}%
  \BibitemOpen
  \bibfield  {author} {\bibinfo {author} {\bibfnamefont {G.}~\bibnamefont
  {Turinici}},\ } {\bibfield
   {journal} {\bibinfo  {journal} {Phys. Rev. A}\ }\textbf {\bibinfo {volume}
  {100}},\ \bibinfo {pages} {053403} (\bibinfo {year} {2019})}\BibitemShut
  {NoStop}%
\bibitem{vandamme2017}%
  \BibitemOpen
  \bibfield  {author} {\bibinfo {author} {\bibfnamefont {L.}~\bibnamefont
  {Van~Damme}}, \bibinfo {author} {\bibfnamefont {Q.}~\bibnamefont {Ansel}},
  \bibinfo {author} {\bibfnamefont {S.~J.}\ \bibnamefont {Glaser}},\ and\
  \bibinfo {author} {\bibfnamefont {D.}~\bibnamefont {Sugny}},\ } {\bibfield  {journal} {\bibinfo
  {journal} {Phys. Rev. A}\ }\textbf {\bibinfo {volume} {95}},\ \bibinfo
  {pages} {063403} (\bibinfo {year} {2017})}\BibitemShut {NoStop}%
\bibitem{Ruschhaupt_2012}%
  \BibitemOpen
  \bibfield  {author} {\bibinfo {author} {\bibfnamefont {A.}~\bibnamefont
  {Ruschhaupt}}, \bibinfo {author} {\bibfnamefont {X.}~\bibnamefont {Chen}},
  \bibinfo {author} {\bibfnamefont {D.}~\bibnamefont {Alonso}},\ and\ \bibinfo
  {author} {\bibfnamefont {J.~G.}\ \bibnamefont {Muga}},\ } {\bibfield  {journal}
  {\bibinfo  {journal} {New Journal of Physics}\ }\textbf {\bibinfo {volume}
  {14}},\ \bibinfo {pages} {093040} (\bibinfo {year} {2012})}\BibitemShut
  {NoStop}%
\bibitem{vandamme2017_2}%
  \BibitemOpen
  \bibfield  {author} {\bibinfo {author} {\bibfnamefont {L.}~\bibnamefont
  {Van-Damme}}, \bibinfo {author} {\bibfnamefont {D.}~\bibnamefont {Schraft}},
  \bibinfo {author} {\bibfnamefont {G.~T.}\ \bibnamefont {Genov}}, \bibinfo
  {author} {\bibfnamefont {D.}~\bibnamefont {Sugny}}, \bibinfo {author}
  {\bibfnamefont {T.}~\bibnamefont {Halfmann}},\ and\ \bibinfo {author}
  {\bibfnamefont {S.}~\bibnamefont {Gu\'erin}},\ } {\bibfield  {journal} {\bibinfo
  {journal} {Phys. Rev. A}\ }\textbf {\bibinfo {volume} {96}},\ \bibinfo
  {pages} {022309} (\bibinfo {year} {2017})}\BibitemShut {NoStop}%
\bibitem{daems2013}%
  \BibitemOpen
  \bibfield  {author} {\bibinfo {author} {\bibfnamefont {D.}~\bibnamefont
  {Daems}}, \bibinfo {author} {\bibfnamefont {A.}~\bibnamefont {Ruschhaupt}},
  \bibinfo {author} {\bibfnamefont {D.}~\bibnamefont {Sugny}},\ and\ \bibinfo
  {author} {\bibfnamefont {S.}~\bibnamefont {Gu\'erin}},\ } {\bibfield  {journal}
  {\bibinfo  {journal} {Phys. Rev. Lett.}\ }\textbf {\bibinfo {volume} {111}},\
  \bibinfo {pages} {050404} (\bibinfo {year} {2013})}\BibitemShut {NoStop}%
\bibitem{genov2014}%
  \BibitemOpen
  \bibfield  {author} {\bibinfo {author} {\bibfnamefont {G.~T.}\ \bibnamefont
  {Genov}}, \bibinfo {author} {\bibfnamefont {D.}~\bibnamefont {Schraft}},
  \bibinfo {author} {\bibfnamefont {T.}~\bibnamefont {Halfmann}},\ and\
  \bibinfo {author} {\bibfnamefont {N.~V.}\ \bibnamefont {Vitanov}},\ } {\bibfield  {journal}
  {\bibinfo  {journal} {Phys. Rev. Lett.}\ }\textbf {\bibinfo {volume} {113}},\
  \bibinfo {pages} {043001} (\bibinfo {year} {2014})}\BibitemShut {NoStop}%
\bibitem{barnes1}%
  \BibitemOpen
  \bibfield  {author} {\bibinfo {author} {\bibfnamefont {D.}~\bibnamefont
  {Buterakos}}, \bibinfo {author} {\bibfnamefont {S.}~\bibnamefont
  {Das~Sarma}},\ and\ \bibinfo {author} {\bibfnamefont {E.}~\bibnamefont
  {Barnes}},\ } {\bibfield
  {journal} {\bibinfo  {journal} {PRX Quantum}\ }\textbf {\bibinfo {volume}
  {2}},\ \bibinfo {pages} {010341} (\bibinfo {year} {2021})}\BibitemShut
  {NoStop}%
\bibitem{barnes2}%
  \BibitemOpen
  \bibfield  {author} {\bibinfo {author} {\bibfnamefont {J.}~\bibnamefont
  {Zeng}}\ and\ \bibinfo {author} {\bibfnamefont {E.}~\bibnamefont {Barnes}},\
  } {\bibfield  {journal}
  {\bibinfo  {journal} {Phys. Rev. A}\ }\textbf {\bibinfo {volume} {98}},\
  \bibinfo {pages} {012301} (\bibinfo {year} {2018})}\BibitemShut {NoStop}%
\bibitem{barnes3}%
  \BibitemOpen
  \bibfield  {author} {\bibinfo {author} {\bibfnamefont {J.}~\bibnamefont
  {Zeng}}, \bibinfo {author} {\bibfnamefont {C.~H.}\ \bibnamefont {Yang}},
  \bibinfo {author} {\bibfnamefont {A.~S.}\ \bibnamefont {Dzurak}},\ and\
  \bibinfo {author} {\bibfnamefont {E.}~\bibnamefont {Barnes}},\ } {\bibfield  {journal} {\bibinfo
  {journal} {Phys. Rev. A}\ }\textbf {\bibinfo {volume} {99}},\ \bibinfo
  {pages} {052321} (\bibinfo {year} {2019})}\BibitemShut {NoStop}%
\bibitem{madaysalomon}%
  \BibitemOpen
  \bibfield  {author} {\bibinfo {author} {\bibfnamefont {Y.}~\bibnamefont
  {Maday}}\ and\ \bibinfo {author} {\bibfnamefont {J.}~\bibnamefont
  {Salomon}},\ }in\ \href@noop {} {\emph {\bibinfo {booktitle} {{Proceedings of
  the 48th IEEE Conference on Decision and Control, 2009, Held jointly whit the
  28th Chinese Control Conference (CDC/CCC 2009)}}}},\ \bibinfo {series and
  number} {{IEEE Conference on Decision and Control}}\ (\bibinfo {year}
  {{2009}})\ pp.\ \bibinfo {pages} {{375--379}}\BibitemShut {NoStop}%
\bibitem{BCS2021}%
  \BibitemOpen
  \bibfield  {author} {\bibinfo {author} {\bibfnamefont {S.}~\bibnamefont
  {Buchwald}}, \bibinfo {author} {\bibfnamefont {G.}~\bibnamefont
  {Ciaramella}},\ and\ \bibinfo {author} {\bibfnamefont {J.}~\bibnamefont
  {Samonon}},\ }\href@noop {} {\bibfield  {journal} {\bibinfo  {journal} {to
  appear in SIAM J. Control Optim.}\ } (\bibinfo {year} {2021})}\BibitemShut
  {NoStop}%
\bibitem{conolly:1986}%
  \BibitemOpen
  \bibfield  {author} {\bibinfo {author} {\bibfnamefont {S.}~\bibnamefont
  {Conolly}}, \bibinfo {author} {\bibfnamefont {D.}~\bibnamefont {Nishimura}},\
  and\ \bibinfo {author} {\bibfnamefont {A.}~\bibnamefont {Macovski}},\
  }\href@noop {} {\bibfield  {journal} {\bibinfo  {journal} {IEEE Trans. Med.
  Imaging}\ }\textbf {\bibinfo {volume} {5}},\ \bibinfo {pages} {106} (\bibinfo
  {year} {1986})}\BibitemShut {NoStop}%
\bibitem{lapertprl}%
  \BibitemOpen
  \bibfield  {author} {\bibinfo {author} {\bibfnamefont {M.}~\bibnamefont
  {Lapert}}, \bibinfo {author} {\bibfnamefont {Y.}~\bibnamefont {Zhang}},
  \bibinfo {author} {\bibfnamefont {M.}~\bibnamefont {Braun}}, \bibinfo
  {author} {\bibfnamefont {S.~J.}\ \bibnamefont {Glaser}},\ and\ \bibinfo
  {author} {\bibfnamefont {D.}~\bibnamefont {Sugny}},\ } {\bibfield  {journal}
  {\bibinfo  {journal} {Phys. Rev. Lett.}\ }\textbf {\bibinfo {volume} {104}},\
  \bibinfo {pages} {083001} (\bibinfo {year} {2010})}\BibitemShut {NoStop}%
\bibitem{Levitt08}%
  \BibitemOpen
  \bibfield  {author} {\bibinfo {author} {\bibfnamefont {M.}~\bibnamefont
  {Levitt}},\ }\href@noop {} {\emph {\bibinfo {title} {Spin Dynamics: Basics of
  Nuclear Magnetic Resonance}}}\ (\bibinfo  {publisher} {Wiley},\ \bibinfo
  {address} {Chichester, UK},\ \bibinfo {year} {2008})\BibitemShut {NoStop}%
\bibitem{skinner:2005}%
  \BibitemOpen
  \bibfield  {author} {\bibinfo {author} {\bibfnamefont {T.~E.}\ \bibnamefont
  {Skinner}}, \bibinfo {author} {\bibfnamefont {T.~O.}\ \bibnamefont {Reiss}},
  \bibinfo {author} {\bibfnamefont {B.}~\bibnamefont {Luy}}, \bibinfo {author}
  {\bibfnamefont {N.}~\bibnamefont {Khaneja}},\ and\ \bibinfo {author}
  {\bibfnamefont {S.~J.}\ \bibnamefont {Glaser}},\ }\href@noop {} {\bibfield
  {journal} {\bibinfo  {journal} {J. Magn. Reson.}\ }\textbf {\bibinfo {volume}
  {172}},\ \bibinfo {pages} {17} (\bibinfo {year} {2005})}\BibitemShut
  {NoStop}%
  \bibitem{suppmat} Numerical results are available upon request to the corresponding author
\end{thebibliography}




%
\end{document}